\documentstyle[epsfig]{lamuphys}
\makeatletter
\let\chapter\hid@chapter
\makeatother
\begin{document}
\pagenumbering{arabic}
\title{Discrete and Global Symmetries in Particle Physics}
\author{R. D. Peccei}
\institute{Department of Physics and Astronomy, UCLA, Los Angeles, CA 
90095-1547}
\maketitle

\begin{abstract}
I begin these lectures by examining the transformation properties of quantum
fields under the discrete symmetries of Parity, P, Charge Conjugation, C,
and Time Reversal, T.  With these results in hand, I then show how the
structure of the Standard Model helps explain the conservation/violation
of these symmetries in various sectors of the theory.  This discussion is
also used to give a qualitative proof of the CPT Theorem, and some of the
stringent tests of this theorem in the neutral Kaon sector are reviewed.
In the second part of these lectures, global symmetries are examined.  Here,
after the distinction between Wigner-Weyl and Nambu-Goldstone realizations
of these symmetries is explained, a discussion is given of the various,
approximate or real, global symmetries of the Standard Model.  Particular
attention is paid to the role that chiral anomalies play in altering the
classical symmetry patterns of the Standard Model.  To understand the
differences between anomaly effects in QCD and those in the electroweak
theory, a discussion of the nature of the vacuum structure of gauge theories
is presented.  This naturally raises the issue of the strong CP problem, and
I present a brief discussion of the chiral solution to this problem and of
its ramifications for astrophysics and cosmology.  I also touch briefly on
possible constraints on, and prospects for, having real Nambu-Goldstone 
bosons in nature, concentrating specifically on the simplest example of
Majorons.  I end these lectures by discussing the compatibility of having
global symmetry in the presence of gravitational interactions.  Although
these interactions, in general, produces small corrections, they can alter
significantly the Nambu-Goldstone sector of theories.
\end{abstract}

%\newpage

\section{Discrete Space-Time Symmetries}

Lorentz transformations
\begin{equation}
x^\mu\to x^{\prime\mu} = \Lambda^\mu_{~\nu}x^\nu
\end{equation}
preserve the invariance of the space-time interval
\begin{equation}
x_\mu x^\mu = \vec r^2-c^2t^2 = \vec r^{\prime^2} - c^2t^{\prime^2} =
x^\prime_\mu x^{\prime\mu}~.
\end{equation}
This constrains the matrices $\Lambda^\mu_{~\nu}$ to obey
\begin{equation}
\eta_{\mu\nu} = \Lambda^\lambda_{~\mu}\eta_{\lambda\kappa}\Lambda^\kappa_{~\nu}~,
\end{equation}
where the matrix tensor $\eta_{\mu\nu}$ is the diagonal matrix
\begin{equation}
\eta_{\mu\nu} = 
\left[ \begin{array}{cccc}
-1 & \hfil & \hfil & \hfil \\
\hfil & 1 & \hfil & \hfil \\
\hfil & \hfil & 1 & \hfil \\
\hfil & \hfil & \hfil & 1
\end{array} \right]~.
\end{equation}
The pseudo-orthogonality of the $\Lambda$ matrices detailed in Eq. (3)
\begin{equation}
\eta = \Lambda^T\eta\Lambda
\end{equation}
allows the classification of Lorentz transformations depending on whether
\begin{equation}
{\rm det}~\Lambda = \left\{
\begin{array}{l}
+1 \\ -1
\end{array} \right. ~; ~~~
\Lambda^0_0 = \pm
\sqrt{1+\sum^3_{i=1} (\Lambda^i_{~0})^2} =
\left\{
\begin{array}{l}
\geq +1 \\ \leq -1
\end{array} \right.~.
\end{equation}
As a result, the Lorentz group splits into four distinct pieces
\begin{eqnarray}
L_+^\uparrow:~{\rm det}~\Lambda &=& +1;~\Lambda^0_0 \geq 1 \nonumber \\
L_-^\uparrow:~{\rm det}~\Lambda &=& -1;~\Lambda^0_0 \geq 1 \nonumber \\
L_+^\downarrow:~{\rm det}~\Lambda &=& +1;~\Lambda_0^0 \leq -1 \nonumber \\
L_-^\downarrow:~{\rm det}~\Lambda &=& -1;~\Lambda_0^0 \leq -1~.
\end{eqnarray}

The transformation matrices $\Lambda$ in $L_+^\uparrow$ by themselves form
a sub-group of the Lorentz group: the proper orthochronous Lorentz group.
All other transformations in the Lorentz group can be obtained from
$\Lambda$ in $L^\uparrow_+$ by using two discrete transformations, P and
T, characterized by the matrices:
\begin{equation}
P^\mu_{~\nu} = \left[
\begin{array}{cccc}
+1 & \hfil & \hfil & \hfil \\
\hfil & -1 & \hfil & \hfil \\
\hfil & \hfil & -1 & \hfil \\
\hfil & \hfil & \hfil & -1
\end{array} \right]~;
T^\mu_{~\nu} = \left[
\begin{array}{cccc}
-1 & \hfil & \hfil & \hfil \\
\hfil & +1 & \hfil & \hfil \\
\hfil & \hfil & +1 & \hfil \\
\hfil & \hfil & \hfil & +1
\end{array} \right]
\end{equation}
corresponding to space inversion (Parity) and time reversal.  It is clear
that if $\Lambda\in L^\uparrow_+$, then $P\Lambda\in L^\uparrow_-$;
$PT\Lambda\in L^\downarrow_+$; and $T\Lambda\in L^\downarrow_-$.  Remarkably,
nature is invariant only under the proper orthochronous Lorentz 
transformations.  Parity is violated in the
weak interactions, something which was first suggested by Lee and Yang (\cite{LY})
in 1956 and soon thereafter observed experimentally (\cite{pviol}).  The
detection of the decay of $K^0_L$ into pions by Christenson, Cronin, Fitch
and Turlay  (\cite{CCFT}) in 1964 provided indirect evidence that also time
reversal is not a good symmetry of nature.

One can understand why this is so on the basis of the Standard Model of
electroweak and strong interactions and of the, so called, CPT theorem,
established by Pauli, Schwinger, L\"uders and Zumino (\cite{CPT}).  To appreciate
these facts I will need to sketch how quantum fields behave under the
discrete space-time transformations of P and T, as well as their behavior
under charge conjugation (C) which physically corresponds to reversing
the sign of all charges.  I will begin with parity.

\subsection{Parity}

The Parity transformation properties of the electromagnetic fields follow
directly from classical considerations.\footnote{Henceforth, I shall use
natural units where $c=\hbar=1$.}  The Lorentz force
\begin{equation} 
\vec F = \frac{d\vec p}
{dt} = q(\vec E+\vec v\times\vec B)
\end{equation} 
obviously changes sign under Parity, since
$\vec p\to -\vec p$.\footnote{Since Parity reverses the sign of space
coordinates $\vec r\to -\vec r$, the velocity also changes sign,
$\vec v\to -\vec v$.} 
Hence, it follows that $\vec E$ is odd and $\vec B$
is even under Parity:
\begin{equation}
\vec E(\vec x,t) \stackrel{P}{\longrightarrow} -\vec E(-\vec x,t)~; ~~
\vec B(\vec x,t)\stackrel{P}{\longrightarrow} B(-\vec x,t)~.
\end{equation}

Formally, the transformation above is induced by a Unitary operator
$U(P)$.  This operator takes the vector potential $A^\mu(\vec x,t)$ into
a transformed vector potential $A^\mu(-\vec x,t)$.  In view of Eq. (10),
it is easy to see that
\begin{equation}
U(P) A^\mu(\vec x,t)U(P)^{-1} =
\eta(\mu)A^\mu(-\vec x,t)~,
\end{equation}
where the symbol $\eta(\mu)$ is a useful notational shorthand, with
\begin{equation}
\eta(\mu) = 
\left\{ \begin{array}{ll}
-1 & \mu \not= 0 \\
+1 & \mu = 0~.
\end{array} \right.
\end{equation}

Spin-zero scalar, $S(\vec x,t)$, and pseudoscalar, $P(\vec x,t)$, fields
under parity are, respectively, even and odd.  That is,
\begin{eqnarray}
U(P)S(\vec x,t)U(P)^{-1} &=& S(-\vec x,t) \nonumber \\
U(P)P(\vec x,t)U(P)^{-1} &=& -P(-\vec x,t) ~.
\end{eqnarray}
The behavior of spin-1/2 Dirac fields $\psi(\vec x,t)$ under Parity is
slightly more complex.  However, this behavior can be straightforwardly
deduced from the requirement that the Dirac equation be invariant under this
operation.  One finds that
\begin{equation}
U(P)\psi(\vec x,t)U(P)^{-1} = \eta_{\rm P}\gamma^0\psi(-\vec x,t)~.
\end{equation}
Here $\eta_{\rm P}$ is a phase factor of unit magnitude $(|\eta_{\rm P}|^2=1)$.
Because one is always interested in fermion-antifermion bilinears, the
phase factor $\eta_{\rm P}$ plays no role physically and one can set it to
unity $(\eta_{\rm P} \equiv 1)$ without loss of generality.

Given Eq. (14), it is a straightforward exercise to deduce the Parity
properties of fermion-antifermion bilinears.\footnote{In my
conventions $\{\gamma^\mu,\gamma^\nu\}=-2\eta^{\mu\nu},~\gamma^{0^{\dagger}}
=\gamma^0$ but $\gamma^{i^{\dagger}} = -\gamma^i$, and $\gamma_5 =
i\gamma^0\gamma^1\gamma^2\gamma^3$.}  Since
\begin{equation}
\gamma^0\gamma^0\gamma^0 = \gamma^0~;
\gamma^0\gamma^i\gamma^0 = -\gamma^i~;
\gamma^0\gamma_5\gamma^0 = -\gamma_5~,
\end{equation}
one easily deduces that
\begin{eqnarray}
U(P)\bar\psi(\vec x,t)\psi(\vec x,t)U(P)^{-1} &=&
\bar\psi(-\vec x,t)\psi(-\vec x,t)~~\hbox{(Scalar)} \nonumber \\
U(P)\bar\psi(\vec x,t)i\gamma_5\psi(\vec x,t)U(P)^{-1} &=&
-\bar\psi(-\vec x,t)i\gamma_5\psi(-\vec x,t)~~\hbox{(Pseudoscalar)}
\nonumber \\
U(P)\bar\psi(\vec x,t)\gamma^\mu\psi(\vec x,t)U(P)^{-1} &=&
\eta(\mu)\bar\psi(-\vec x,t)\gamma^\mu\psi(-\vec x,t)~~\hbox{(Vector)}
\nonumber \\
U(P)\bar\psi(\vec x,t)\gamma^\mu\gamma_5\psi(\vec x,t)U(P)^{-1} &=&
-\eta(\mu)\bar\psi(-\vec x,t)\gamma^\mu\gamma_5\psi(-\vec x,t)~~
\hbox{(Pseudovector)}\nonumber \\
&&
\end{eqnarray}
From the above, one sees immediately that the electromagnetic
interaction is parity invariant:
\begin{equation}
W^{\rm em}_{\rm int} =
\int d^4xeA^\mu(x)\bar\psi(x)\gamma_\mu\psi(x)
\stackrel{P}{\longrightarrow} W^{\rm em}_{\rm int}~.
\end{equation}
On the other hand, because Parity transforms fields of a given
{\bf chirality} into each other\footnote{Here $\psi_{\rm L}(x) =
\frac{1}{2}(1-\gamma_5)\psi(x);~\psi_{\rm R}(x) = \frac{1}{2}
(1+\gamma_5)\psi_{\rm R}(x)$.}
\begin{equation}\psi_{\rm L}(\vec x,t)\stackrel{P}{\longrightarrow}
\gamma^0\psi_{\rm R}(-\vec x,t);~
\psi_{\rm R}(\vec x,t)\stackrel{P}{\longrightarrow}
\gamma^0\psi_{\rm L}(-\vec x,t)~,
\end{equation}
it is obvious that the chirally asymmetric weak interactions will
violate parity.  Thus, this sector of the Standard Model is
Parity violating.  The strong interactions, however, are invariant
under Parity.  These interactions are governed by Quantum Chromodynamics 
and in QCD
both the left-handed and right-handed quarks are triplets under
the $SU(3)$ gauge group:
\begin{equation}
q_{\rm L} \sim 3~;~~q_{\rm R} \sim 3~.
\end{equation}
Note the difference here with respect to the weak interactions.
Under the weak $SU(2)$ group of the $SU(2)\times U(1)$ theory,
the left-handed fields $\psi_{\rm L}$ of both quarks and leptons
are doublets, while the right-handed fields $\psi_{\rm R}$ are
singlets
\begin{equation}
\psi_{\rm L} \sim 2~;~~\psi_{\rm R} \sim 1~.
\end{equation}
This is the root cause for the violation of Parity in the weak
interactions.

\subsection{Charge Conjugation}

As I alluded to earlier, the process of charge conjugation is
connected physically with the reversal of the sign of all electric
charges.  For the electromagnetic field, therefore, the charge
conjugation transformation C brings the vector potential
$A^\mu(x)$ into minus itself
\begin{equation}
U(C)A^\mu(x) U(C)^{-1} = -A^\mu(x)~.
\end{equation}
For Dirac fields, since charge conjugation should transform
particles into antiparticles, this operation essentially
corresponds to Hermitian conjugation.  That is, one has
\begin{equation}
U(C)\psi(x) U(C)^{-1} =
\eta_cC\psi^{\dagger}(x)~.
\end{equation}
Here $\eta_c$ is again a phase factor of unit magnitude and,
without loss of generality, one can take $\eta_c\equiv 1$.  The
form of the matrix C can be deduced from the requirement that
the transformation (22) should leave the Dirac equation invariant.
For this to be the case necessitates that
\begin{equation}
C\gamma_\mu^*C^{-1} = -\gamma_\mu~.
\end{equation}
The particular form of C one obtains depends on the form of
the $\gamma$-matrices used.  In the Majorana representation, where
the $\gamma$-matrices are purely imaginary [Majorana:
$\gamma^*_\mu = -\gamma_\mu$] then $C=1$.  On the other hand, in
the Dirac representation [Dirac: $\gamma^0 =
\left[ \begin{array}{cc}
1 & 0 \\ 0 & -1
\end{array} \right]~;~~
\gamma^i = \left[ \begin{array}{cc}
0 & \sigma^i \\ -\sigma^i & 0
\end{array} \right]~$],
then $C=\gamma_2$.  Because of the simplicity of $C$ in the
Majorana representation, in what follows we shall make use of this
representation when dealing with charge conjugation.

Using Eq. (22), it is straightforward to compute the $C$-conjugation properties
of fermion antifermion bilinears.  Let me do this explicitly for the scalar
density $\bar\psi\psi$ and then quote the results for the other bilinears.
One has
\begin{eqnarray}
U(C)\bar\psi(x)\psi(x)U(C)^{-1} &=& 
U(C)\psi_\alpha^{\dagger}(x)(\gamma^0)_{\alpha\beta}\psi_\beta(x) U(C)^{-1}
\nonumber \\
&=&\psi_\alpha(x)(\gamma^0)_{\alpha\beta}\psi^{\dagger}_\beta(x) \nonumber \\
&=&-\psi^{\dagger}_\beta(x)(\gamma^0)_{\alpha\beta}\psi_\alpha(x) \nonumber \\
&=&-\psi^{\dagger}_\beta(x)(\gamma^{0T})_{\beta\alpha}\psi_\alpha(x) \nonumber \\
&=&+\bar\psi(x)\psi(x)~.
\end{eqnarray}
The second line above is the result of using Eq. (22), taking $C=1$ assuming
one is working in the Majorana representation.  The third line above follows
because fermion fields anticommute (apart from an irrelevant infinite piece
which can be subtracted away).  Finally, the last line follows since in the
Majorana representation $\gamma^0$ is an antisymmetric matrix
$(\gamma^{0T} = -\gamma^0)$.

The full set of results for the behavior of fermion-antifermion bilinears under
$C$ is displayed below:
\begin{eqnarray}
&U(C)&\bar\psi(x)\psi(x)U(C)^{-1} = \bar\psi(x)\psi(x)~~\hbox{(Scalar)} 
\nonumber \\
&U(C)&\bar\psi(x)i\gamma_5\psi(x)U(C)^{-1} = 
\bar\psi(x)i\gamma_5\psi(x)~~\hbox{(Pseudoscalar)} \nonumber \\
&U(C)&\bar\psi(x)\gamma^\mu\psi(x)U(C)^{-1} =
-\bar\psi(x)\gamma^\mu\psi(x)~~\hbox{(Vector)} \nonumber \\
&U(C)&\bar\psi(x)\gamma^\mu\gamma_5\psi(x)U(C)^{-1} =
\bar\psi(x)\gamma^\mu\gamma^5\psi(x)~~\hbox{(Pseudovector)}~.
\end{eqnarray}
These results lead to some immediate consequences.  For instance, it follows
that electromagnetic interactions are $C$-invariant.  Using Eqs. (21) and
(25) it follows that
\begin{equation}
W^{\rm em}_{\rm int} = \int d^4xeA^\mu(x)\bar\psi(x)\gamma_\mu\psi(x)
\stackrel{C}{\longrightarrow} W^{\rm em}_{\rm int}~,
\end{equation}
since both $A^\mu$ and the electromagnetic current $\bar\psi\gamma^\mu\psi$
change sign under $C$.

The strong interactions are also invariant under charge conjugation.  This
takes a small discussion, but it is also easy to see.  The principal point
to note is that the $SU(3)$ currents of QCD do not have the same simple
transformation properties as the electromagnetic current, because they involve
the non-trivial $SU(3)$ matrices $\lambda_a$.  Effectively these matrices get
transposed in the bilinears, if one makes a charge conjugation transformation.
That is, one has
\begin{equation}
U(C) \bar q\gamma^\mu\frac{\lambda_a}{2} qU(C)^{-1} =
-\bar q\gamma^\mu\left(\frac{\lambda_a}{2}\right)^T~q~.
\end{equation}
Because $\lambda_1,\lambda_3,\lambda_4,\lambda_6$, and $\lambda_8$ are
symmetric, while $\lambda_2,\lambda_5$, and $\lambda_7$ are antisymmetric,
it follows that
\begin{equation}
J_a^\mu \to -\eta(a)J_a^\mu~,
\end{equation}
where
\begin{equation}
\eta(a) = \left\{
\begin{array}{ll}
+1 & \hbox{for}~a=1,3,4,6~{\rm and}~8 \\
-1 & \hbox{for}~a=2,5~{\rm and}~7
\end{array} \right.
\end{equation}
To guarantee invariance of the quark gluon interaction terms
\begin{equation}
W_{\rm int} = \int d^4xg_3A^\mu_aJ_{\mu a}
\end{equation}
under charge conjugation it is necessary to assume that the charge conjugation
properties of the gluon fields themselves vary according to which component
one is dealing with.  Namely, for invariance of Eq. (30) under $C$ one needs
\begin{equation}
U(C)A_a^\mu(x)U(C)^{-1} = -\eta(a)A_a^\mu(x)~.
\end{equation}

It is easy to check that the above transformation property is precisely what
is needed to have the nonlinear gluon field strengths have well defined $C$-properties.  Recall that
\begin{equation}
G_a^{\mu\nu} = \partial^\mu A_a^\nu-\partial^\nu A_a^\mu +
gf_{abc}A_b^\mu A_c^\nu~.
\end{equation}
Now, for $SU(3)$, the only non-vanishing structure constants $f_{abc}$ are (\cite{Slansky})
\begin{equation}
f_{abc}\not= 0~\hbox{for}~
abc = \{123,147,156,246,257,345,367,458,678\}~.
\end{equation}
One sees that $f_{abc}\not= 0$ only for cases in which there is an odd number of
indices which themselves are odd (i.e. the indices: 2,5, and 7).  This assures
that, indeed, $G^{\mu\nu}_a$ transforms in the same way as $A_a^\mu$ does
under $C$:
\begin{equation}
U(C) G_a^{\mu\nu}(x)U(C)^{-1} = -\eta(a)G_a^{\mu\nu}(x)~.
\end{equation}
This last property then insures that
\begin{equation}
W^{\rm QCD} = \int d^4x\left[-\bar q\left(\gamma^\mu\frac{1}{i} D_\mu + 
m_q\right) q - \frac{1}{4} G_a^{\mu\nu}G_{a\mu\nu}\right]
\stackrel{C}{\longrightarrow} W^{\rm QCD}~.
\end{equation}

The situation is different for the weak interactions since these involve both
vector and pseudovector interactions.  Let us focus, for example, on the
$SU(2)$ current for leptons of the first generation
\begin{equation}
J_i^\mu = (\bar\nu_e~\bar e)_{\rm L}\gamma^\mu\frac{\tau_i}{2}
\left( \begin{array}{c}
\nu_e \\ e 
\end{array} \right)_{\rm L} =
\frac{1}{4}(\bar\nu_e~\bar e)\gamma^\mu(1-\gamma_5)\tau_i 
\left( \begin{array}{c}
\nu_e \\ e
\end{array} \right)~.
\end{equation}
This current transforms differently in its vector and pseudovector pieces
as well as in its 1, 3 and 2 components:
\begin{eqnarray}
U(C)J^\mu_{1,3}U(C)^{-1} &=& -\frac{1}{4}(\bar\nu_e~\bar e)\gamma^\mu
(1+\gamma_5)\tau_{1,3} \left(
\begin{array}{c}
\nu_e \\ e
\end{array} \right) \nonumber \\
U(C) J^\mu_2 U(C)^{-1} &=& +\frac{1}{4} (\bar\nu_e~\bar e)\gamma^\mu
(1+\gamma_5)\tau_2 \left( 
\begin{array}{c}
\nu_e \\ e
\end{array} \right)~.
\end{eqnarray}
The difference in behavior in the 1,3 and 2 components is absorbed by
postulating the following $C$-transformation properties for the
$W_i^\mu$ fields.\footnote{These transformation properties guarantee that
$F_i^{\mu\nu}$ and $W_i^\mu$ have the same $C$-properties.}
\begin{equation}
U(C) W_i^\mu(x) U(C)^{-1} = -\eta(i)W_i^\mu(x)~,
\end{equation}
with
\begin{equation}
\eta(i) = \left\{
\begin{array}{ll}
+1 & i=1,3 \\
-1 & i=2 
\end{array} \right.
\end{equation}
Note that these properties are what one might expect since they imply that
\begin{equation}
W_\pm^\mu = \frac{1}{\sqrt{2}} (W_1^\mu \mp iW_2^\mu)
\stackrel{C}{\longrightarrow} -W_\mp^\mu~.
\end{equation}
However, even so, the simultaneous presence of vector and pseudovector pieces
in the currents which enter the weak interactions force one to conclude
that
\begin{equation}
W_{\rm weak~interactions} \stackrel{C}{\not\!\longrightarrow} 
W_{\rm weak~interactions}~,
\end{equation}
as is observed experimentally.

\subsection{Time Reversal}

Classically, $T$-invariance corresponds to the fact that the equations of motion
describing a particle going from $A$ to $B$ along some path also allow, as a
permitted motion, the time reversed motion.  That is, a motion where the
particle follows the same path, but is now going from $B$ to $A$.  Clearly,
in this time reversed motion all momenta are reflected, but the coordinates
remain the same.  So, classically, under a $T$-transformation
\begin{equation}
\vec p \stackrel{T}{\longrightarrow} -\vec p~; ~~
\vec F = \frac{d\vec p}{dt} \stackrel{T}{\longrightarrow} \vec F~.
\end{equation}

Quantum mechanically, the interchange of initial and final states is implemented
by having the operator $U(T)$, corresponding to time reversal, be an
{\bf anti-unitary} operator (\cite{Wigner}), with
\begin{equation}
U(T) = V(T)K~.
\end{equation}
In the above, $V(T)$ is a unitary operator while $K$ complex conjugates any
$c$-number quantity it acts on.  The operation of complex conjugation as part
of $U(T)$ is what renders this operator anti-unitary.  The need for complex
conjugation, in connection with time reversal, is already seen at the level
of the Schr\"odinger equation.  From
\begin{equation}
i\frac{\partial}{\partial t} \psi(\vec x,t) = H\psi(\vec x,t)
\end{equation}
one deduces that $\psi^*(\vec x,-t)$ obeys the equation
\begin{equation}
i\frac{\partial}{\partial t} \psi^*(\vec x,-t) = H^*\psi^*(\vec x,-t)~.
\end{equation}
So, provided that the Hamiltonian is real $(H^*=H)$, then one sees that
$\psi^*(\vec x,-t)$ is also a solution of the Schr\"odinger equation.  
Therefore, in quantum mechanics, complex conjugation of the wave function (along
with the reality of the Hamiltonian) accompanies the reversal in the
direction of time.

The association of complex conjugation with time reversal effectively
interchanges incoming and outgoing states (\cite{Low})
\begin{equation}
\langle U(T)\phi|U(T)\psi\rangle = \langle\psi|\phi\rangle~.
\end{equation}
Thus, if $T$ is a good symmetry of the theory, one relates processes to their
time reversed process (e.g. the decay $A\to BC$ to the formation of $A$ from
the coalescence of $B$ and $C$, $BC\to A$).  More precisely, if time reversal
is a good symmetry, then one relates the $S$-matrix element $S_{fi}$ to that
for $S_{\tilde i\tilde f}$, where the states, 
$\tilde i,\tilde f$ have all the momentum
directions $\{\vec p\}$ reversed in comparison to the states $i,f$.  That is
\begin{equation}
S_{fi} =~_{\rm out}\langle f|i\rangle_{\rm in} 
=~_{\rm in}\langle U(T)i|U(T)f\rangle_{\rm out}
=~_{\rm out}\langle\tilde i|\tilde f\rangle_{\rm in} = S_{\tilde i\tilde f}~.
\end{equation}
The next to last step above is only valid if time reversal is a good symmetry
of the theory, since in this case it follows that
\begin{equation}
U(T)|f\rangle_{\rm out} = |\tilde f\rangle_{\rm in}~; ~~
U(T)|i\rangle_{\rm in} = |\tilde i\rangle_{\rm out}~.
\end{equation}

I should add a comment here about the issue of the reality of the Hamiltonian
needed for time reversal to hold at the Schr\"odinger equation level.  This
is not quite the case when spin is involved and is the reason for the
possible additional operator $V(T)$ in the definition of $U(T)$ in Eq. (43).
More correctly, in general, what is needed is that
\begin{equation}
V(T)H^*V(T)^{-1} = H~.
\end{equation}
When there is no spin $V(T)$ is just the unit matrix, but with spin its
presence allows for $T$-invariance.  The simplest example of this is provided
by the ordinary spin-orbit interaction of atomic physics
\begin{equation}
H_{\rm s-o} = \lambda\vec\sigma\cdot\vec L~,
\end{equation}
with $\lambda$ some real constant.  Since $\vec L = \vec r\times
\frac{1}{i}\vec\nabla$, it follows that
\begin{equation}
H^*_{\rm s-o} = \lambda\vec\sigma^*\cdot\vec L^* =
-\lambda\vec\sigma^*\cdot\vec L~,
\end{equation}
which is not the same as Eq. (50) because $\sigma^*_2 = -\sigma_2$ but
$\sigma^*_{1,3} = \sigma_{1,3}$.  However, since $\sigma_2\vec\sigma^*\sigma_2
=-\vec\sigma$, using $V(T) = \sigma_2$ guarantees that
\begin{equation}
V(T) H^*_{\rm s-o}V(T)^{-1} = H_{\rm s-o}~,
\end{equation}
reflecting physically that, indeed, time reversal not only changes
$\vec L\to -\vec L$, but also, effectively, $\vec\sigma\to -\vec\sigma$.

In field theory, it is again straightforward to deduce what is the effect of
a time-reversal transformation on the electromagnetic fields by focusing on
what happens classically.  Since the Lorentz force is invariant under $T$
\begin{equation}
\vec F = \frac{d\vec p}{dt} = q(\vec E + \vec v\times\vec B) 
\stackrel{T}{\longrightarrow} \vec F~,
\end{equation}
it follows that $\vec E$ is even and $\vec B$ is odd under time-reversal.
In terms of the vector potential, therefore, one has
\begin{equation}
U(T) A^\mu(\vec x,t)U(T)^{-1} = \eta(\mu) A^\mu(\vec x,-t)~.
\end{equation}

For spin-1/2 fields one can deduce the transformation properties of
$\psi(\vec x,t)$ under $T$-transformations by again asking that the action of
$U(T)$ on $\psi(\vec x,t)$ produce another solution of the Dirac equation.
Writing
\begin{equation}
U(T)\psi(\vec x,t)U(T)^{-1} = \eta_TT\psi(\vec x,-t)~,
\end{equation}
with $\eta_T$ a phase of unit magnitude (which we shall take, without loss of
generality,
 to be unity, $\eta_T\equiv 1$), and remembering that $U(T)$ complex
conjugates all $c$-numbers, one finds that for invariance
of the Dirac equation the matrix $T$ must
obey
\begin{eqnarray}
T\gamma^{0*}T^{-1}&=&\gamma^0 \nonumber \\
T\gamma^{i*}T^{-1}&=&-\gamma^i~.
\end{eqnarray}
As was the case for the charge conjugation matrix $C$, the form of the matrix
$T$ also depends on which representation of the $\gamma$-matrices one uses.
In the convenient Majorana representation, where $\gamma^{\mu *}=-\gamma^\mu$,
one finds that
\begin{equation}
T = \gamma^0\gamma_5~.
\end{equation}

Armed with Eqs. (55) and (57), a simple calculation then produces the
following transformation properties for the familiar fermion-antifermion
bilinears:\footnote{In deducing Eq. (58), care must be taken to remember that
$U(T)$ complex conjugates $c$-numbers.}
\begin{eqnarray}
&U(T)&\bar\psi(\vec x,t)\psi(\vec x,t)U(T)^{-1} =
\bar\psi(\vec x,-t)\psi(\vec x,-t)~~\hbox{(Scalar)} \nonumber \\
&U(T)&\bar\psi(\vec x,t)i\gamma_5\psi(\vec x,t)U(T)^{-1} =
-\bar\psi(\vec x,-t)i\gamma_5\psi(\vec x,-t)~~\hbox{(Pseudoscalar)} \nonumber \\
&U(T)&\bar\psi(\vec x,t)\gamma^\mu\psi(\vec x,t)U(T)^{-1} =
\eta(\mu)\bar\psi(\vec x,-t)\gamma^\mu\psi(\vec x,-t)~~\hbox{(Vector)}
\nonumber \\
&U(T)&\bar\psi(\vec x,t)\gamma^\mu\gamma_5\psi(\vec x,t)U(T)^{-1} =
\eta(\mu)\bar\psi(\vec x,-t)\gamma^\mu\gamma_5\psi(\vec x,-t)~~
\hbox{(Pseudovector)}\nonumber \\
&&
\end{eqnarray}

It is obvious from the above and Eq. (54), as well from the reality of the
electromagnetic coupling constant $e$, that the electromagnetic interactions
are $T$-invariant
\begin{equation}
W^{\rm em}_{\rm int} = \int d^4xeA^\mu(x)\bar\psi(x)
\gamma_\mu\psi(x) \stackrel{T}{\longrightarrow} W^{\rm em}_{\rm int}~.
\end{equation}

It is easy to check also that the gauge interactions in both QCD and the
$SU(2)\times U(1)$ electroweak theory are also $T$-invariant, provided one
properly defines how the gauge fields transform.  Since for $SU(3)$ only
$\lambda_2,\lambda_5$ and $\lambda_7$ are imaginary, and for $SU(2)$ only
$\sigma_2$ is imaginary, it is easy to check that the desired
$T$-transformation properties are:\footnote{Of course, the gauge coupling
constants, just like $e$, are real.}
\begin{eqnarray}
U(T)A^\mu_a(\vec x,t)U(T)^{-1}&=&
\eta(\mu)\eta(a)A_a^\mu(\vec x,-t)~~(SU(3)) \nonumber \\
U(T)W_i^\mu(\vec x,t)U(T)^{-1} &=&
\eta(\mu)\eta(i)W_i^\mu(\vec x,-t)~~(SU(2)) \nonumber \\
U(T)Y^\mu(\vec x,t)U(T)^{-1} &=&
\eta(\mu)Y^\mu(\vec x,-t)~~~~~~~~~(U(1))~.
\end{eqnarray}
Note that in contrast to $C$, $T$-transformations affect vector and 
pseudovector currents in the same way.  Thus, using (58) and (60), it follows
immediately that
\begin{equation}
W^{\rm SM}_{\rm gauge~interactions} \stackrel{T}{\longrightarrow}
W^{\rm SM}_{\rm gauge~interactions}~.
\end{equation}

The Standard Model can have, however, $T$-violating interactions in the
electroweak sector involving the scalar Higgs field.  The couplings of the Higgs
field, in contrast to the gauge couplings, do not need to be real.  These
complex couplings then provide the possibility of having $T$-violating
interactions.  I examine this point in the simplest case where one has only
one complex Higgs doublet
\begin{equation}
\Phi = \left(
\begin{array}{c}
\phi^0 \\ \phi^-
\end{array} \right)
\end{equation}
in the theory.  The scalar Higgs self-interactions, which trigger the
breakdown of $SU(2)\times U(1)$, only involve real coefficients since one must
require the Higgs potential to be Hermitian.  That is
\begin{equation}
V = \lambda \left(
\Phi^{\dagger}\Phi - \frac{v^2}{2}\right)^2
= V^{\dagger}
\end{equation}
implies that both $\lambda$ and $v$ are real parameters.

The Yukawa interactions of $\Phi$ with the quark fields, however, can have
complex coefficients.\footnote{I concentrate here only on the quark sector,
because if one does not introduce right-handed neutrinos in the theory---so that
neutrinos are effectively massless---then all the phases in the Yukawa couplings
in the lepton sector can be rotated away.}
With $i,j$ being family indices, one can write, in general, these interactions
as
\begin{equation}
{\cal{L}}_{\rm Yukawa} = - \Gamma^u_{ij}(\bar u,\bar d)_{{\rm L}i}
\Phi u_{{\rm R}j} - \Gamma^d_{ij}(\bar u,\bar d)_{{\rm L}i}
\tilde\Phi d_{{\rm R}j} + \hbox{h.c.}~.
\end{equation}
Here $\tilde\Phi = i\sigma_2\Phi^*$ and the coefficient matrices
$\Gamma^u_{ij}~\Gamma^d_{ij}$ are arbitrary complex matrices.  After the
electroweak interactions are spontaneously broken $(SU(2)\times U(1)
\to U(1)_{\rm em})$, effectively all that remains of the doublet field $\Phi$
is one scalar excitation---the Higgs boson $H$---and the vacuum expectation
value $v$:
\begin{equation}
\Phi\to\frac{1}{\sqrt{2}} \left(
\begin{array}{c}
v+H \\ 0
\end{array} \right)
\end{equation}
Thus the Yukawa interactions (64) generate mass terms for the charge 2/3
and charge -1/3 quarks
\begin{equation}
M_{ij}^{u,d} = \frac{1}{\sqrt{2}}\Gamma_{ij}^{u,d}v~.
\end{equation}
As is well known, these mass matrices can be diagonalized by a bi-unitary
transformation
\begin{equation}
(U_{\rm L}^{u,d})^{\dagger}M^{u,d}U_{\rm R}^{u,d} =
{\cal{M}}^{u,d}~.
\end{equation}

The diagonal matrices ${\cal{M}}^{u,d}$ have real eigenvalues $m_i$,
corresponding to the physical quark masses.  Further, the bi-unitary transformations on the quark fields diagonalizes the Yukawa coupling matrices,
since $M$ and $\Gamma$ are linearly related.  Whence, all that remains of the
Yukawa sector after these transformations is the simple interaction
\begin{equation}
{\cal{L}}^{\rm eff}_{\rm Yukawa} = -\sum_{i} m_i\bar q_i(x)q_i(x)
\left[1 + \frac{H(x)}{v}\right]~.
\end{equation}
Provided $H(\vec x,t)$ has the canonical $T$-transformation one expects for a
scalar field,
\begin{equation}
U(T)H(\vec x,t)U(T)^{-1} = H(\vec x,-t)~.
\end{equation}
Eq. (68) is a $T$-conserving interaction also.  Nevertheless, the complex
nature of the original Yukawa couplings does end up by producing some
$T$-violating interactions.

It is easy to understand this last point.  The bi-unitary transformations
performed on the quarks to diagonalize the quark mass matrices alter the form of
the charged current weak interactions.  Before these transformations, these
interactions had the form
\begin{equation}
{\cal{L}}^{\rm cc} = \frac{e}{2\sqrt{2}~\sin\theta_W}
\left[W_+^\mu J^0_{-\mu} + W_-^\mu J_{+\mu}^0\right]~,
\end{equation}
with
\begin{equation}
J_{-\mu}^0 = (\bar u_1,\bar u_2,\bar u_3)\gamma_\mu(1-\gamma_5)~ {\bf 1}~
\left( 
\begin{array}{c}
d_1 \\ d_2 \\ d_3
\end{array} \right)
\end{equation}
and
\begin{equation}
J^0_{+\mu} = (J^0_{-\mu})^{\dagger}~.
\end{equation}
Clearly, this interaction is $T$-invariant.  However, after the bi-unitary
transformation on the quark fields to diagonalize $M$ [Eq. (67)], the
charged current $J^0_{-\mu}$ is altered to
\begin{equation}
J_{-\mu} = (\bar u,\bar c,\bar t)\gamma_\mu(1-\gamma_5)
{\bf V}_{\rm CKM} \left(
\begin{array}{c}
d \\ s \\ b
\end{array} \right)~,
\end{equation}
where the Cabibbo-Kobayashi-Maskawa quark mixing matrix (\cite{CKM})
\begin{equation}
{\bf V}_{\rm CKM} = U_{\rm L}^{u^{\dagger}}U_{\rm L}^d
\end{equation}
is a unitary matrix, since $U_{\rm L}^u$ and $U_{\rm L}^d$ are.  Because,
in general, ${\bf V}_{\rm CKM}$  is complex, its presence in the currents
$J^\mu_-$ (and $J^\mu_+$) can lead to $T$-violation.

For three families of quarks and leptons, as we apparently have, it is not
difficult to show that the matrix ${\bf V}_{\rm CKM}$ has only one physical phase,
$\delta$.  All the other phases can be rotated away through further harmless
redefinitions of the quark fields.  If $\delta\not= 0$, then the charged
current weak interactions are not $T$-invariant
\begin{equation}
{\cal{L}}^{\rm cc}(\vec x,t) =
\frac{e}{2\sqrt{2}~\sin\theta_{\rm W}} \left[W_+^\mu J_{-\mu} + W^\mu_-J_{+\mu}
\right]
\stackrel{T}{\not\!\longrightarrow} {\cal{L}}^{\rm cc}(\vec x,-t)
\end{equation}
and the standard model can give rise to observable manifestations of 
$T$-violation.  We return to this point in more detail in the next subsection,
after we discuss the CPT theorem.

\subsection{The CPT Theorem}

If nature is described by a local Lorentz invariant field theory, where there
is the usual connection between spin and statistics, then one can prove a deep
theorem, now known as the CPT Theorem (\cite{CPT}).  Namely, in these circumstances,
one can show that the action of the theory is {\bf always invariant} under the
combined application of a $C$-, a $P$-, and a $T$-transformation.  That is
\begin{equation}
W\stackrel{CPT}{\longrightarrow} W~.
\end{equation}
I will not attempt here to establish the CPT theorem with rigor.  The
interested reader can turn, for example, to the erudite manuscript of Streater
and Wightman (\cite{ST}) for this.  Rather, I want to show why and how the CPT
Theorem works, based on the preceding discussion of the C, P, and T
transformation properties of quantum fields.

To get started, let us look at the effect of a CPT transformation on the
electromagnetic interactions.  Using Eqs. (11), (16), (21), (25), (54), and
(58), one has
\begin{eqnarray}
A^\mu(\vec x,t) &\stackrel{CPT}{\longrightarrow}& [-1][\eta(\mu)]
[\eta(\mu)]A^\mu(-\vec x,-t) = -A^\mu(-\vec x,-t) \nonumber \\
J^\mu_{\rm em}(x,t) &=& \bar\psi(\vec x,t)\gamma^\mu\psi(\vec x,t) \nonumber \\
&\stackrel{CPT}{\longrightarrow}& [-1][\eta(\mu)][\eta(\mu)]
J^\mu_{\rm em}(-\vec x,-t) = -J^\mu_{\rm em}(-\vec x,-t)~.
\end{eqnarray}
Obviously, therefore, under a CPT transformation
\begin{equation}
W^{\rm em}_{\rm int} = \int d^4xeA^\mu(x)
J_\mu^{\rm em}(x)\stackrel{CPT}{\longrightarrow}W^{\rm em}_{\rm int}~.
\end{equation}
This, however, is a trivial case, since $W^{\rm em}_{\rm int}$ was
{\bf separately} invariant under C-, P-, and T-transformations!

CPT invariance, if it is a general property, must hold also when there is
violation of the individual symmetries.  A more significant test is provided
by the electroweak theory.  There, for example, both C and P are violated in
the neutral current interactions, while T and CPT are conserved.  Let us
check this.  The action for the neutral current interactions is given by
\begin{equation}
W_{\rm int}^{\rm NC} = \frac{e}{2\cos\theta_W\sin\theta_W}
\int d^4xZ_\mu J^\mu_{\rm NC}~.
\end{equation}
The neutral current
\begin{equation}
J^\mu_{\rm NC} = 2[J^\mu_3 - \sin^2\theta_WJ^\mu_{\rm em}] =
V^\mu + A^\mu
\end{equation}
contains both vector and pseudovector pieces, since these latter components are present
in the $SU(2)$ current $J^\mu_3$.  Parity and Charge Conjugation are violated
in Eq. (79) because the vector and pseudovector currents transform in opposite
ways under each of these transformations.  That is, one has, under Parity
\begin{eqnarray}
&Z^\mu(\vec x,t)\stackrel{P}{\longrightarrow}\eta(\mu)Z^\mu(-\vec x,t)&;~~
V^\mu(\vec x,t)\stackrel{P}{\longrightarrow}\eta(\mu)V^\mu(-\vec x,t);
\nonumber \\
&A^\mu(\vec x,t)\stackrel{P}{\longrightarrow}-\eta(\mu)A^\mu(-\vec x,t)&
\end{eqnarray}
while, under Charge Conjugation,
\begin{equation}
Z^\mu(\vec x,t)\stackrel{C}{\longrightarrow}-Z^\mu(\vec x,t);~
V^\mu(\vec x,t)\stackrel{C}{\longrightarrow}-V^\mu(\vec x,t);
A^\mu(\vec x,t)\stackrel{C}{\longrightarrow} A^\mu(\vec x,t)
\end{equation}
On the other hand, T is conserved by Eq. (79), since under time reversal
\begin{eqnarray}
&Z^\mu(\vec x,t)\stackrel{T}{\longrightarrow}\eta(\mu)Z^\mu(\vec x,-t);&~
V^\mu(\vec x,t)\stackrel{T}{\longrightarrow}\eta(\mu)V^\mu(\vec x,-t);
\nonumber \\
&A^\mu(\vec x,t)\stackrel{T}{\longrightarrow}\eta(\mu)\vec A(\vec x,-t)&.
\end{eqnarray}
Using the above three equations, it is easy to see that the neutral current
interactions conserve CPT.  One has
\begin{eqnarray}
&Z^\mu(x,t)\stackrel{CPT}{\longrightarrow}-Z^\mu(-\vec x,-t)&;~
V^\mu(\vec x,t)\stackrel{CPT}{\longrightarrow}-V^\mu(-\vec x,-t);
\nonumber \\
&A^\mu(\vec x,t)\stackrel{CPT}{\longrightarrow}-A^\mu(-\vec x,-t)&.
\end{eqnarray}
From the above, it is also clear that CP and T are {\bf equivalent}
transformations for the neutral current action
\begin{equation}
W^{\rm NC}_{\rm int} \stackrel{CP}{\longrightarrow} W^{\rm NC}_{\rm int}
\stackrel{T}{\longrightarrow}W^{\rm NC}_{\rm int}~.
\end{equation}

The equivalence between a T-transformation and a CP-transformation also holds
when both of these potential symmetries are violated.  Hence, even in this
case, the combined CPT-transformation is indeed an invariance of the action.
This is the essence of the CPT Theorem.  To appreciate this point let me
examine, specifically, the T-violating charged current interaction between the
$u$ and $b$ quarks, typified by the complex CKM matrix element $V_{ub}$.\footnote{One can pick phase conventions where $V_{ub}$ is real.  In
this case, however, other pieces in the charged current Lagrangian give rise
to T-violation.  The final result for physically measured parameters must be 
phase-convention independent.  I focus here on the $V_{ub}$ term for 
definitiveness, since in the standard convention for the CKM matrix (\cite{CKM})
$V_{ub}$ is complex and its phase is precisely $-\delta$.}  One has
\begin{equation}
W^{\rm cc}_{ub} = \frac{e}{2\sqrt{2}~\sin\theta_W} \int d^4x
\left\{V_{ub}W^\mu_+\bar u\gamma_\mu(1-\gamma_5)b+V^*_{ub}W^\mu_-
\bar b\gamma_\mu(1-\gamma_5)u\right\}~,
\end{equation}
where
\begin{equation}
W_\pm^\mu = \frac{1}{\sqrt{2}}(W_1^\mu\mp iW_2^\mu)~.
\end{equation}
Because under $T$
\begin{equation}
W_1^\mu(\vec x,t)\stackrel{T}{\longrightarrow}\eta(\mu)W_1^\mu(\vec x,-t);~
W_2^\mu(\vec x,t)\stackrel{T}{\longrightarrow}-\eta(\mu)W_2^\mu(\vec x,-t)
\end{equation}
and remembering the $i$ factor in Eq. (87), it follows that
\begin{equation}
W_\pm^\mu(\vec x,t)\stackrel{T}{\longrightarrow}\eta(\mu)W_\pm^\mu(\vec x,-t)~.
\end{equation}
On the other hand, under $T$, the $u-b$ currents behave as
\begin{eqnarray}
\bar u(\vec x,t)\gamma_\mu(1-\gamma_5)b(\vec x,t)&\stackrel{T}{\longrightarrow}&
\eta(\mu)\bar u(\vec x,-t)\gamma_\mu(1-\gamma_5)b(\vec x,-t) \nonumber \\
\bar b(\vec x,t)\gamma_\mu(1-\gamma_5)u(\vec x,t)&\stackrel{T}{\longrightarrow}&
\eta(\mu)\bar b(\vec x,-t)\gamma_\mu(1-\gamma_5)u(\vec x,-t)~.
\end{eqnarray}
Hence, one sees, indeed, that the action $W^{\rm cc}_{ub}$ is not T-invariant
\begin{equation}
W^{\rm cc}_{ub}\stackrel{T}{\longrightarrow}\tilde W^{\rm cc}_{ub} =
\frac{e}{2\sqrt{2}\sin\theta_W} \int d^4x\left\{V^*_{ub}W_+^\mu\bar u\gamma_\mu
(1-\gamma_5)b + V_{ub}W_-^\mu\bar b\gamma_\mu(1-\gamma_5)u\right\}.
\end{equation}

The behavior of the various ingredients in $W^{\rm cc}_{ub}$ under CP is
individually different than it is under T.  For instance, one has
\begin{equation}
W_1^\mu(\vec x,t)\stackrel{CP}{\longrightarrow}-\eta(\mu)W_1^\mu(-\vec x,t);~
W_2^\mu(\vec x,t)\stackrel{CP}{\longrightarrow}\eta(\mu)W_2^\mu(-\vec x,t)~.
\end{equation}
Hence, since one also does not complex conjugate the $i$ in $W^\mu_\pm$ in this
case, one has
\begin{equation}
W_\pm^\mu(\vec x,t)\stackrel{CP}{\longrightarrow}-\eta(\mu)W_\mp(-\vec x,t)~.
\end{equation}
Similarly, one finds, that under CP the $u-b$ currents transform as
\begin{eqnarray}
\bar u(\vec x,t)\gamma_\mu(1-\gamma_5)b(\vec x,t)&\stackrel{CP}{\longrightarrow}&
-\eta(\mu)\bar b(-\vec x,t)\gamma_\mu(1-\gamma_5)u(-\vec x,t) \nonumber \\
\bar b(\vec x,t)\gamma_\mu(1-\gamma_5)u(\vec x,t)&\stackrel{CP}{\longrightarrow}&
-\eta(\mu)\bar u(-\vec x,t)\gamma_\mu(1-\gamma_5)b(-\vec x,t)~.
\end{eqnarray}
The net effect, however, on $W^{\rm cc}_{ub}$ is the same as that of a
T-transformation.  One finds
\begin{equation}
W^{\rm cc}_{ub}\stackrel{CP}{\longrightarrow} \tilde W^{\rm cc}_{ub} =
\frac{e}{2\sqrt{2}~\sin\theta_W} \int d^4x
\left\{V_{ub}W_-^\mu\bar b\gamma_\mu(1-\gamma_5)u+V^*_{ub}W_+^\mu
\bar u\gamma_\mu(1-\gamma_5)b\right\}~.
\end{equation}

One can extract from this example the underlying reason why the CPT theorem
holds.  It results really from a combination of the needed Hermiticity of the
Lagrangian and the complementary role that T and CP play on the operators
and $c$-numbers that enter in the Lagrangian.  Hermiticity means that a given
term in the Lagrangian, containing some operator $O(x)$ and some $c$-number
$a$, has the form
\begin{equation}
{\cal{L}}(x) = aO(x) + a^*O^{\dagger}(x)~.
\end{equation}
Under T, the operator is unchanged (except for replacing $t$ by $-t$), but the
$c$-number is complex conjugated
\begin{equation}
O(\vec x,t)\stackrel{T}{\longrightarrow}O(\vec x,-t)~; ~~
a\stackrel{T}{\longrightarrow}a^*~.
\end{equation}
Under CP, on the other hand, the operator $O$ gets essentially replaced by its
Hermitian adjoint, but the $c$-number $a$ stays the same:
\begin{equation}
O(\vec x,t)\stackrel{CP}{\longrightarrow} O^{\dagger}(-\vec x,t)~; ~~
a\stackrel{CP}{\longrightarrow} a~.
\end{equation}
Combining the operations of T and CP changes, effectively, the first term in
Eq. (96) into the second term and {\it vice versa}
\begin{equation}
{\cal{L}} = aO(x) + a^*O^{\dagger}(x)\stackrel{CPT}{\longrightarrow}
{\cal{L}}(-x) = a^*O^{\dagger}(-x) + aO(-x)
\end{equation}
leaving the action invariant
\begin{equation}
W = \int d^4x{\cal{L}}(x)\stackrel{CPT}{\longrightarrow} W~.
\end{equation}

\subsection{CP and CPT Tests in the Neutral Kaon Complex}

The $K^0\sim \bar sd$ and $\bar K^0\sim s\bar d$ states provide an excellent laboratory
to test CP and CPT.  These states are unstable, decaying into particles with
no strangeness through a first-order weak process.  In addition, second order
weak processes, giving rise to the transition $\bar sd\leftrightarrow s\bar d$,
allow the $K^0$ to mix with the $\bar K^0$.  The quantum mechanical evolution of
this two-state system leads to the physical eigenstates $K^0_L$ and 
$K^0_{S}$, characterized by their, respective, long and short lifetimes.

The physical eigenstates $K^0_L$ and $K^0_S$ are obtained by diagonalizing
the $2\times 2$ effective Hamiltonian
\begin{equation}
H_{\rm eff} = M - \frac{i}{2}\Gamma~.
\end{equation}
Here $M$ and $\Gamma$ are Hermitian matrices describing the mass mixing and
decay properties of the neutral Kaon complex.  If CPT is a good symmetry of
nature, then the diagonal matrix elements of $M$ and $\Gamma$ are equal, since
this symmetry changes effectively $K^0$ into $\bar K^0$.
\begin{equation}
M_{11} = M_{22}~; ~~\Gamma_{11} = \Gamma_{22}~~
\hbox{[CPT Conservation]}~.
\end{equation}
CP conservation, on the other hand, guarantees the reality of the mass and
decay matrices.  It provides therefore a constraint on the off-diagonal matrix
elements of $M$ and $\Gamma$.  Namely:
\begin{equation}
M_{12} = M^*_{12}~; ~~ \Gamma_{12} = \Gamma^*_{12}~~
\hbox{[CP Conservation]}~.
\end{equation}

If one does not impose the above constraints of CPT and CP conservation on
$M$ and $\Gamma$, the eigenstates of the Schr\"odinger equation
\begin{equation}
H_{\rm eff} \left(
\begin{array}{c}
|K^0\rangle \\ |\bar K^0\rangle
\end{array} \right)
= i\frac{\partial}{\partial t}
\left(
\begin{array}{c}
|K^0\rangle \\ |\bar K^0\rangle
\end{array} \right)
\end{equation}
are linear superpositions of the $|K^0\rangle$ and $|\bar K^0\rangle$ states,
involving parameters $\delta_K$ and $\epsilon_K$ which reflect the breaking
of these symmetries.  The physical $|K_L^0\rangle$ and $|K^0_S\rangle$
eigenstates have the standard time evolution
\begin{equation}
|K_{L,S}(t)\rangle = \exp[-im_{L,S}t]\exp\left[-\frac{1}{2}\Gamma_{L,S}t\right]
|K_{L,S}(0)\rangle~,
\end{equation}
characterized by the mass and width of these particles.  The states
$|K_{L,S}(0)\rangle$ involve the following superposition of
the $|K^0\rangle$ and $|\bar K^0\rangle$ states:
\begin{eqnarray}
|K_L(0)\rangle &=& \frac{1}{\sqrt{2}}\left\{(1+\epsilon_K+\delta_K)|K^0\rangle
+ (1-\epsilon_K-\delta_K)|\bar K^0\rangle\right\} \nonumber \\
|K_S(0)\rangle &=& \frac{1}{\sqrt{2}}\left\{(1+\epsilon_K-\delta_K)|K^0\rangle
-(1-\epsilon_K+\delta_K)|\bar K^0\rangle\right\}~.
\end{eqnarray}
In the above
\begin{eqnarray}
\epsilon_K &=& e^{i\phi_{SW}}\left[\frac{-{\rm Im}~M_{12}+\frac{i}{2}~{\rm Im}~
\Gamma_{12}}{\sqrt{2}~\Delta m}\right] \nonumber \\
\delta_K &=& ie^{i\phi_{SW}}\left[\frac{(M_{11}-M_{22})-\frac{i}{2}
(\Gamma_{11}-\Gamma_{22})}{2\sqrt{2}~\Delta m}\right]~,
\end{eqnarray}
where
\begin{equation}
\phi_{SW} = \tan^{-1}\frac{2\Delta m}{\Gamma_S-\Gamma_L}~;~~
\Delta m = m_L-m_S~.
\end{equation}
Experimentally, one finds (\cite{PDG})
\begin{equation}
\phi_{SW} = (43.49\pm 0.08)^o~;~~
\Delta m = (3.491\pm 0.009)\times 10^{-12}~{\rm MeV}~.
\end{equation}
Note that $\epsilon_K=0$, if CP is conserved and $\delta_K=0$, if CPT is
conserved.  Only if both $\epsilon_K$ and $\delta_K$ vanish are the eigenstates
$|K^0_L\rangle$ and $|K^0_S\rangle$ CP eigenstates.  If both these symmetries
hold then
\begin{equation}
CP|K^0_{L,S}\rangle = \mp|K^0_{L,S}\rangle~~~
\hbox{[CP, CPT Conservation]}~.
\end{equation}

What is measured experimentally are the CP violating ratios of the
amplitude of the $K_L$ and $K_S$ to go into two pions
\begin{eqnarray}
\eta_{+-} &=& \frac{A(K_L\to \pi^+\pi^-)}{A(K_S\to\pi^+\pi^-)} =
|\eta_{+-}|e^{i\phi_{+-}} = \epsilon + \epsilon^\prime \nonumber \\
\eta_{00} &=& \frac{A(K_L\to\pi^0\pi^0)}{A(K_S\to \pi^0\pi^0)} =
|\eta_{00}|e^{i\phi_{00}} = \epsilon-2\epsilon^\prime~.
\end{eqnarray}
Experimentally, one finds that $\eta_{+-}\simeq \eta_{00}$ (so 
$\epsilon\gg \epsilon^\prime)$,
with (\cite{PDG})
\begin{equation}
|\eta_{+-}| = (2.285 \pm 0.019)\times 10^{-3}~;~~
\phi_{+-} = (43.7\pm 0.6)^o~.
\end{equation}

Neglecting the contribution of the widths compared to the masses, which is a
very good approximation, one finds that the parameter $\epsilon$ above is
simply (\cite{BCDP})
\begin{equation}
\epsilon\simeq \epsilon_K-\delta_K\simeq e^{i\phi_{SW}}\left[
\frac{-{\rm Im}~M_{12}}{\sqrt{2}~\Delta m}\right] + ie^{i\phi_{SW}} \left[
\frac{M_{22}-M_{11}}{2\sqrt{2}~\Delta m}\right]~.
\end{equation}
Note that the CPT violating contribution in the above is $90^o$ 
{\bf out of phase}
from the CP violating contribution.  Because $\phi_{SW} = (43.49\pm
0.08)^o$ is consistent with $\phi_{+-}=(43.7\pm 0.6)^o$, one deduces
immediately that the non-zero value for $\eta_{+-}$ observed is mostly a signal
of CP-violation $[{\rm Im}~M_{12}\not= 0]$ rather than of CPT violation
[$M_{11}\not= M_{22}$].

If one neglects altogether the possibility that there is any CPT violation
in the neutral Kaon decay amplitudes---something one would eventually need to
check---then one can write approximately
\begin{equation}
M_{22}-M_{11}\simeq |\eta_{+-}|2\sqrt{2}~\Delta m\tan(\phi_{+-}-\phi_{SW})~.
\end{equation}
This equation, given the values of the experimental parameters involved, provides
a spectacularly strong bound on CPT violation, because the $K_L-K_S$ mass
difference $\Delta m$ is so small.  One finds, at the 90\% CL,
\begin{equation}
\left|\frac{m_{\bar K^0}-m_{K^0}}{m_{K^0}}\right| < 9\times 10^{-19}~,
\end{equation}
which is an incredibly stringent test of CPT.

Experiments at the just completed Frascati Phi Factory will be able to directly
measure $\delta_K$, without further assumptions, to an accuracy similar to the
present
accuracy for $\epsilon$.  This will be accomplished by studying the difference in
relative time decay patterns of the doubly semileptonic decays of the
$K_LK_S$ states produced in the $\Phi$ decay.  If one studies the relative
time dependence of the process $\Phi\to K_LK_S\to \pi^-e^+\nu_e(t_1)\pi^+e^-
\bar\nu_e(t_2)$, then one can show that the pattern at large $\Delta t =
t_1-t_2$ is sensitive to ${\rm Re}~\delta_K$, while the pattern at small
$\Delta t$ is sensitive to ${\rm Im}~\delta_K$ (\cite{BCDP}).

\section{Continuous Global Symmetries}

In the Standard Model there are a variety of global symmetries, both exact and
approximate.  Some of these symmetries are manifest [Wigner-Weyl realized],
while others are spontaneously broken [Nambu-Goldstone realized].  I wish
here to examine these matters in some detail.

An important distinction exists for a continuous global symmetry depending on
whether or not the vacuum state respects the symmetry.  Let us denote the
global symmetry group for the theory by $G$.  This group, in
general, will have generators $g_i$ which obey an algebra
\begin{equation}
[g_i,g_j] = ic_{ijk}g_k~,
\end{equation}
where $c_{ijk}$ are the structure constants for the group.  If the
generators $g_i$, for all $i$, annihilate the vacuum
\begin{equation}
g_i|0\rangle = 0~,
\end{equation}
then the symmetry group is realized in a Wigner-Weyl way, with degenerate
multiplets of states in the spectrum (\cite{WW}).  If, on the other hand, for
some generators $g_i$
\begin{equation}
g_i|0\rangle \not= 0
\end{equation}
then the symmetry group $G$ is spontaneously broken to a subgroup
$H~(G\to H)$ and $n = {\rm dim}~G/H$ massless scalars appear in the spectrum
of the theory.  This is the Nambu-Goldstone realization of the symmetry
$G$ and the massless scalars are known as Nambu-Goldstone 
bosons (\cite{NG}).

Physically, approximate global symmetries are easy to understand.  These
symmetries result from being able to neglect dynamically certain parameters
in the theory.  A well known example is provided by Quantum Chromodynamics
(QCD).  The Lagrangian of QCD
\begin{equation}
{\cal{L}}_{\rm QCD} = -\sum_i \bar q_i\left(\gamma^\mu\frac{1}{i} D_\mu +
m_i\right) q_i - \frac{1}{4} G_a^{\mu\nu}G_{a\mu\nu}
\end{equation}
has an approximate global symmetry, connected to the fact that the
lightest quark masses $m_u$ and $m_d$ are much smaller than the dynamical scale
of the theory, $\Lambda_{\rm QCD}$.\footnote{The strange quark mass $m_s \sim
\Lambda_{\rm QCD}$ may also be neglected in some circumstances, leading to a
larger $SU(3)\times SU(3)$ global symmetry.}  Neglecting the light
quark masses, one sees that the QCD Lagrangian is invariant under a large global
symmetry transformation
\begin{equation}
{\cal{L}}_{\rm QCD} \stackrel{U(n_f)_{\rm L}\times U(n_f)_{\rm R}}{\longrightarrow} {\cal{L}}_{\rm QCD}~,
\end{equation}
where $n_f$ is the number of flavors whose masses are neglected.  Under this
group of transformations the $n_f$ light quarks go into each other.  For
example, for $n_f=2$, neglecting $m_u$ and $m_d$ in the QCD Lagrangian
allows the symmetry transformation
\begin{equation}
\left( \begin{array}{c}
u \\ d
\end{array} \right)_{\rm L}
\to e^{ia_{i{\rm L}}T_i} \left(
\begin{array}{c}
u \\ d
\end{array} \right)_{\rm L}~; ~~
\left( \begin{array}{c}
u \\ d
\end{array} \right)_{\rm R}
\to e^{ia_{i{\rm R}}T_i} \left(
\begin{array}{c}
u \\ d
\end{array} \right)_{\rm R}~;
\end{equation}
where $T_i=(\tau_i,1)$.

The global $U(2)_{\rm L}\times U(2)_{\rm R}$ approximate symmetry of QCD,
arising from the fact that $m_u,m_d \ll \Lambda_{\rm QCD}$, is actually only
a symmetry at the classical level.  At the quantum level, there is an 
Adler-Bell-Jackiw (\cite{ABJ}) anomaly in a $U(1)_{\rm R-L}$ subgroup 
of this symmetry and the
real approximate global symmetry of QCD is reduced to
\begin{equation}
G=SU(2)_{\rm R+L}\times SU(2)_{\rm R-L}\times U(1)_{\rm R+L} \equiv
SU(2)_V\times SU(2)_A\times U(1)_B~.
\end{equation}
Only $SU(2)_V$ and $U(1)_B$, however, are manifest symmetries of nature.  The
$SU(2)_A$ symmetry is spontaneously broken by the formation of $u$ and $d$
quark condensates, due to the QCD dynamics (see,  for example, \cite{chiral})
\begin{equation}
\langle\bar uu\rangle = \langle\bar dd\rangle\not= 0~.
\end{equation}
The manifest $SU(2)_V$ symmetry, is the well-known isospin symmetry of the
strong interactions (\cite{Heisenberg}), leading to the approximate nucleon
$N = (p,n)$ and pion $\pi = (\pi^\pm,\pi^0)$ multiplets.  $U(1)_B$
corresponds to baryon number and its existence as a good symmetry guarantees
that nucleons and antinucleons have the same mass.  The spontaneously broken
$SU(2)_A$ symmetry leads to the appearance of three Nambu-Goldstone bosons,
which are identified as the pions.  Indeed, one can show that (see, for example, \cite{Peccei})
\begin{equation}
m_\pi^2 \to 0~~\hbox{as}~~m_{u,d}\to 0~.
\end{equation}

Although $SU(2)_V\times SU(2)_A$ are only {\bf approximate} symmetries of QCD,
valid of we neglect $m_u$ and $m_d$ in the QCD Lagrangian, $U(1)_B$ is
actually an exact global symmetry of the theory corresponding to the
transformation
\begin{equation}
q_i\to\exp\left[\frac{i}{3}\alpha_B\right]~q_i~.
\end{equation}
This transformation, since it affects all quarks equally, is also clearly
a symmetry of the electroweak theory.  Indeed, since all interactions always
involve $q-\bar q$ pairs, it follows immediately that
\begin{equation}
{\cal{L}}_{\rm SM} \stackrel{U(1)_B}{\longrightarrow} {\cal{L}}_{\rm SM}~,
\end{equation}
with the associated conserved current being given by
\begin{equation}
J^\mu_B = \frac{1}{3} \sum_i \bar q_i\gamma^\mu q_i~.
\end{equation}

Precisely the same argument can be made for leptons, since again all 
interactions in the Standard Model always involve a lepton-antilepton pair.
Whence, one has
\begin{equation}
{\cal{L}}_{\rm SM} \stackrel{U(1)_L}{\longrightarrow}{\cal{L}}_{\rm SM}~,
\end{equation}
with
\begin{equation}
J_{\rm L}^\mu = \sum_i \bar\ell_i\gamma^\mu\ell_i
\end{equation}
being the corresponding conserved current.  

At the quantum level, however, it
turns out that neither $U(1)_{\rm L}$ or $U(1)_{\rm B}$ are good symmetries,
because of the chiral nature of the weak interactions.  Because the 
left-handed fields under the $SU(2)\times U(1)$ Standard Model group behave
differently than the right-handed fields, effectively in the electroweak
theory both $J^\mu_{\rm B}$ and $J^\mu_{\rm L}$ feel corresponding ABJ
anomalies (\cite{'tHooft}).  As we shall see, the breaking of $U(1)_{\rm B}$
and $U(1)_{\rm L}$ by these anomalies is the same.  Hence, in the electroweak
theory, at the quantum level, there remains only one true global quantum
symmetry, $U(1)_{\rm B-L}$:
\begin{equation}
{\cal{L}}_{\rm SM}\stackrel{U(1)_{\rm B-L}}{\longrightarrow} {\cal{L}}_{\rm SM}~.
\end{equation}
We shall soon discuss these matters in some detail.  However, before doing so,
let me remark that the electroweak theory has actually a larger set of global
symmetries if the neutrino masses vanish $(m_{\nu_i}=0)$.\footnote{Theoretically,
this is simply achieved by not including any right-handed neutrino fields
$\nu_{{\rm R}_i}$ in the theory.}  In this case, each {\bf individual}
lepton number $(L_e,L_\mu$ and $L_\tau)$ is separately conserved at the
classical level, while, say, $3L_e-B,~3L_\mu-B,~3L_\tau - B$ are conserved
at the quantum level.

If one includes right-handed neutrinos in the standard model, so that
$m_{\nu_i} \not= 0$, then one expects in general neutrino mixing, much as in the
quark case.  One knows, however, experimentally that neutrino masses, if they
exist at all are very light (\cite{PDG})---typically with masses in the eV range.
With such light neutrino masses, effectively the Standard Model produces
extremely small lepton flavor violations.  For instance, one knows experimentally
that (\cite{PDG})
\begin{equation}
BR(\mu\to e\gamma) < 5\times 10^{-11}~.
\end{equation}
Such a transition can occur at the one-loop level in the SM, but its ratio is
extremely suppressed due to the tiny neutrino masses (\cite{Fritzsch}). 
Typically, one finds
\begin{equation}
BR(\mu\to e\gamma) \sim \frac{\alpha G_F\sin\theta_\nu(m^2_{\nu_i}-
m^2_{\nu_2})}{M^2_W} \sim 10^{-24}~.
\end{equation}
Here $\theta_\nu$ is a neutrino mixing angle and the numerical result 
corresponds to taking $\sin\theta_\nu\sim 10^{-1}$ and $\Delta m^2_\nu\sim
({\rm eV})^2$.

\subsection{Chiral Anomalies}

The existence of chiral anomalies (\cite{ABJ}) has important consequences for the
Standard Model.  Anomalies, as we shall see, alter the classical global
symmetry structure of the model.  In addition, they bring into play the gauge
field strength structure
\begin{equation}
F_a^{\mu\nu}\tilde F_{a\mu\nu} = \frac{1}{2}
\epsilon^{\mu\nu\alpha\beta} F_{a\alpha\beta}F_{a\mu\nu}~.
\end{equation}
This structure is C even, but is both P and T odd.  Hence, it can provide
additional sources of CP violation.  In the Standard Model, it does so through
the, so-called, $\bar\theta$-term effective interaction
\begin{equation}
{\cal{L}}_{\rm CP~viol.} = \bar\theta\frac{\alpha_3}{8\pi}
G_a^{\mu\nu}\tilde G_{a\mu\nu}~,
\end{equation}
where $G_a^{\mu\nu}$ is the gluon field strength for QCD and $\alpha_3$ is
the corresponding (squared) coupling constant $[\alpha_3 = g_3^2/4\pi]$.

For pedagogical reasons, it is important to sketch the {\it raison d'etre}
for chiral anomalies.  This is done best in the simple example provided by a
theory which has a single fermion field $\psi$ and a $U(1)_V\times U(1)_A$
global symmetry.  In such a theory, at the classical (Lagrangian) level there
are two conserved currents
\begin{equation}
J_V^\mu = \bar\psi\gamma^\mu\psi~~\hbox{with}~~
\partial_\mu J_V^\mu = 0
\end{equation}
and
\begin{equation}
J_A^\mu = \bar\psi\gamma^\mu\gamma_5\psi~~\hbox{with}~~
\partial_\mu J_A^\mu = 2m\bar\psi i\gamma_5\psi
\stackrel{m\to 0}{\longrightarrow} 0~.
\end{equation}
That is, the chiral $U(1)_A$ symmetry obtains if the fermion $\psi$ is massless.
At the quantum level, however, it is not possible to preserve {\bf both}
the conservation laws for $J_A^\mu$ and $J_V^\mu$.  This is the origin of the
chiral anomaly  (\cite{ABJ}).

\begin{figure}
\begin{center}
\epsfig{file=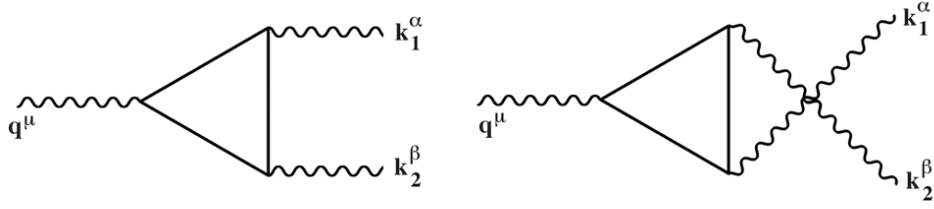,width=5in}
\end{center}
\caption[]{Triangle graphs contributing to the AVV anomaly}
\end{figure}

More specifically, the source of the anomaly is the singular behavior of the
triangle graph (shown in Fig. 1) involving one axial current $J_A^\mu$ and two
vector currents
$J_V^\mu$.  The individual graphs in Fig. 1 are each logarithmic
divergent.  However, their sum is finite.  One can write the Green's function
for two vector currents $J_V^\mu$ and an axial current as (\cite{Adler})
\begin{equation}
T^{\mu\alpha\beta} = F(q^2,k_1^2,k_2^2)P^{\mu\alpha\beta}(k_1,k_2)~.
\end{equation}
The pseudotensor $P^{\mu\alpha\beta}(k_1,k_2)$ by Bose symmetry obeys
\begin{equation}
P^{\mu\alpha\beta}(k_1,k_2) = P^{\mu\beta\alpha}(k_2,k_1)~.
\end{equation}
Further, the conservation of the vector currents imposes the constraints
\begin{equation}
k_{1\alpha}P^{\mu\alpha\beta}(k_1,k_2) = k_{2\beta}P^{\mu\alpha\beta}(k_1,k_2)
= 0~.
\end{equation}
The above equations imply a unique structure for the pseudotensor
$P^{\mu\alpha\beta}(k_1,k_2)$, namely
\begin{equation}
P^{\mu\alpha\beta}(k_1,k_2) = \epsilon^{\alpha\beta\rho\sigma}k_{1\rho}
k_{2\sigma}q^\mu~.
\end{equation}
Because of the momentum factors in $P^{\mu\alpha\beta}(k_1,k_2)$, it follows
that the invariant function $F(q^2,k_1^2,k_2^2)$ is indeed finite.

Given the above, imagine regularizing the triangle graphs in Fig. 1 via a
Pauli-Villars regularization, to make each of the individual graphs finite (\cite{Adler}).  Denoting the graphs in Fig. 1, respectively, by
$t^{\mu\alpha\beta}(k_1,k_2)$ and $t^{\mu\beta\alpha}(k_2,k_1)$
this procedure yields for $T^{\mu\alpha\beta}$ the expression
\begin{eqnarray}
T^{\mu\alpha\beta} &=& \epsilon^{\alpha\beta\rho\sigma}k_{1\rho}k_{2\sigma}
q^\mu F(q^2,k_1^2,k_2^2) \nonumber \\
&=&\left[t^{\mu\alpha\beta}(k_1,k_2)\left|_m -
t^{\mu\alpha\beta}(k_1,k_2)\right|_M\right] \nonumber \\
& & \mbox{}+\left[t^{\mu\beta\alpha}(k_2,k_1)\left|_m -
t^{\mu\beta\alpha}(k_2,k_1)\right|_M\right]~.
\end{eqnarray}
Here $M$ is the Pauli-Villars regularization mass.  Taking the divergence of
the above and setting the fermion mass $m\to 0$ yields the expression
\begin{equation}
q_\mu T^{\mu\alpha\beta} = -2iMP^{\alpha\beta}(M)~.
\end{equation}
Here the pseudoscalar structure $P^{\alpha\beta}(M)$ involves similar graphs
to those in Fig. 1, except that the axial vertex is proportional to
$\gamma_5$ and not $\gamma^\mu\gamma_5$.

Because the function $F(q^2,k^2_1,k^2_2)$ is finite, one knows that the
Pauli-Villars regularization is really irrelevant and that one can therefore
let $M\to\infty$.  By straightforward calculation (\cite{Adler}) one finds that
\begin{equation}
\lim_{M\to\infty} -2iMP^{\alpha\beta}(M) = \frac{i}{2\pi^2}
\epsilon^{\alpha\beta\rho\sigma}k_{1\rho}k_{2\sigma}~.
\end{equation}
Hence, one deduces the Adler-Bell-Jackiw anomalous divergence equation  (\cite{ABJ})
\begin{equation}
q_\mu T^{\mu\alpha\beta} = \frac{i}{2\pi^2}
\epsilon^{\alpha\beta\rho\sigma}k_{1\rho}k_{2\sigma}~.
\end{equation}

The anomalous Ward identity for $T^{\mu\alpha\beta}$ above can be interpreted
in terms of an effective violation of the conservation equation for the
axial current $J_A^\mu$.  Because the $U(1)_V$ gauge bosons- ``photons"-couple
to $J_V^\alpha$ and $J_V^\beta$, it is easy to show that Eq. (144) is 
equivalent to the anomalous divergence equation
\begin{equation}
\partial_\mu J_A^\mu = \frac{e^2}{8\pi^2} F_{\alpha\beta}\tilde F^{\alpha\beta} = \frac{\alpha}{2\pi} F_{\alpha\beta}
\tilde F^{\alpha\beta}~,
\end{equation}
where $e$ is the $U(1)_V$ coupling constant.  The above is the famous
Adler-Bell-Jackiw chiral anomaly (\cite{ABJ}).

The above result, whose derivation we sketched for the
$U(1)_V\times U(1)_A$ theory, can easily be generalized to the case where the
fields in the current $J_A^\mu$ carry some non-Abelian charge.  In this case
the fermions in the anomalous triangle graphs carry some non-Abelian index and
the graph, instead of simply involving $e^2$, now contains a factor of
\begin{equation}
g^2~{\rm Tr}~\frac{\lambda_a}{2}\frac{\lambda_a}{2} =
\frac{1}{2} g^2\delta_{ab}~.
\end{equation}
Here $g$ is the coupling constant associated to the non-Abelian group and 
$\lambda_a/2$ is the appropriate generator matrix for the fermion fields,
assuming they transform according to the fundamental representation of the
non-Abelian group.  It follows, therefore, that in the
non-Abelian case the chiral anomaly (145) is replaced by
\begin{equation}
\partial_\mu J_A^\mu = \frac{g^2}{16\pi^2}F_a^{\alpha\beta}
\tilde F_{a\alpha\beta} = \frac{\alpha^2_g}{4\pi}
F_a^{\alpha\beta}\tilde F_{a\alpha\beta}~,
\end{equation}
where $F_a^{\alpha\beta}$ are the field strengths for the non-Abelian
gauge bosons.

One can use the above results to analyze the Baryon (B) and Lepton (L)
number currents in the Standard Model (\cite{'tHooft}).  These currents, as we
mentioned earlier, are conserved at the Lagrangian level.  Decomposing these
currents into chiral components, one has
\begin{eqnarray}
J_{\rm B}^\mu &=& \frac{1}{3}\sum_i \bar q_i
\gamma^\mu q_i = \frac{1}{3}\sum_i
(\bar q_{i{\rm L}}\gamma^\mu q_{i{\rm L}} + \bar q_{i{\rm R}}\gamma^\mu
q_{i{\rm R}}) \nonumber \\
J_{\rm L}^\mu &=& \sum_i \bar\ell_i\gamma^\mu\ell_i = \sum_i
(\bar\ell_{i{\rm L}}\gamma^\mu\ell_{i{\rm L}}+\bar\ell_{i{\rm R}}\gamma^\mu
\ell_{i{\rm R}})~.
\end{eqnarray}
Because the quarks and leptons interact with the $SU(2)\times U(1)$ 
electroweak fields the divergence of $J_{\rm B}^\mu$ and $J_{\rm L}^\mu$
will not vanish, as a result of the chiral anomalies.  A straightforward
computation of the relevant triangle graphs gives
\begin{equation}
\partial_\mu J_{\rm B}^\mu = -\frac{\alpha_2}{8\pi} N_g W_i^{\mu\nu}
\tilde W_{i\mu\nu} + \frac{\alpha_1}{8\pi} N_g \left(\frac{4}{9} +
\frac{1}{9} - \frac{1}{18}\right) Y^{\alpha\beta}\tilde Y_{\alpha\beta}
\end{equation}
and
\begin{equation}
\partial_\mu J_{\rm L}^\mu = -\frac{\alpha_2N_g}{8\pi}
W_i^{\mu\nu}\tilde W_{i\mu\nu} + \frac{\alpha_1}{8\pi} N_g
\left(1-\frac{1}{2}\right) Y^{\alpha\beta}\tilde Y_{\alpha\beta}~.
\end{equation}
In the above, $N_g$ is the number of generations.  The various numbers in
front of the contributions involving the $U(1)$ gauge bosons contain the
squares of the appropriate hypercharges, multiplied by the corresponding number
of states [e.g. $u_{\rm R}$ contributes a factor of 4/9, while the doublet
$(u,d)_{\rm L}$ contributes a factor of $2\times 1/36$].  Note that for the
Baryon number current and for the Lepton number current, not only the $SU(2)$ but
also the $U(1)$ factors are the same [(4/9 + 1/9-1/18) = (1-1/2) = 1/2].
It follows therefore that, as advertized, the total fermion number B+L is
broken at the quantum level, but B-L is conserved:
\begin{eqnarray}
\partial_\mu J^\mu_{\rm B+L} &=& \frac{\alpha_1^2}{8\pi} N_g Y^{\alpha\beta}
\tilde Y_{\alpha\beta} - \frac{\alpha^2_2}{4\pi} N_g W_i^{\alpha\beta}
\tilde W_{i\alpha\beta} \nonumber \\
\partial_\mu J^\mu_{\rm B-L} &=& 0~.
\end{eqnarray}

A similar situation obtains in QCD.  In the limit as $m_u,m_d\to 0$, this
theory has a global symmetry at the classical level of $SU(2)_V\times
SU(2)_A\times U(1)_V\times U(1)_A$.  However, the $U(1)_A$ current
\begin{equation}
J_5^\mu = \frac{1}{2}[\bar u\gamma^\mu\gamma_5u + \bar d\gamma^\mu\gamma_5d]
\end{equation}
has a chiral anomaly, since the quarks carry color and interact with the
gluons.  Taking into account the contribution of both the $u$ and $d$ quarks
in the triangle graph, one finds
\begin{equation}
\partial_\mu J_5^\mu = \frac{\alpha_3^2}{4\pi} G_a^{\alpha\beta}
\tilde G_{a\alpha\beta}~.
\end{equation}

The violation of the (B+L)-current in the electroweak theory and of the
$U(1)_A$ current in QCD, codified by Eqs. (151) and (153), have a similar
aspect.  Nevertheless, these quantum corrections are quite different physically
in their import.  As we shall see, the current $J_5^\mu$ is really badly broken
by the above quantum QCD effects.  As a result, 
as we mentioned earlier, the classical $U(1)_A$ 
symmetry is never a good (approximate) symmetry of the strong interactions.
In contrast, $J^\mu_{\rm B+L}$ is extraordinarily weakly broken by the quantum
corrections, except in the early Universe where temperature-dependent effects
enhance these contributions.  Thus, at zero temperature, the total fermion
number (B+L) is essentially conserved.

Physically, these two results are what is needed.  The formation of
$u$ and $d$-quark condensates
\begin{equation}
\langle\bar uu\rangle = \langle\bar dd\rangle \not= 0
\end{equation}
in QCD clearly breaks both the $SU(2)_A$ and $U(1)_A$ symmetries spontaneously.
If $U(1)_A$ were really a symmetry, one would expect to have an associated
Nambu-Goldstone boson---the $\eta$---with similar properties to the
$SU(2)_A$ Nambu-Goldstone bosons---the $\pi$ mesons.  Although these states are
supposed to be massless when the respective global symmetries are exact,
both states should get similar masses once one includes quark mass terms for
the $u$ and $d$ quarks (\cite{Weinberg}).  However, experimentally, one finds
$m^2_\eta \gg m^2_\pi$ and one concludes that $U(1)_A$ cannot really be a true
symmetry of QCD.  Thus the strong breaking of $J_A^\mu$ by the anomaly is a
welcome result.

In contrast, for the electroweak theory it is important that the anomalous
breaking of (B+L) should not physically lead to large effects, since one has
very strong experimental bounds on baryon number violation.  For instance the
B-violating decay $p\to e^+\pi^0$ has a bound (\cite{PDG})
\begin{equation}
\tau(p\to e^+\pi^0) > 5.5\times 10^{32}~\hbox{years}~.
\end{equation}
To undersand why the anomaly contribution in Eq. (153) connected to the
$U(1)_A$ current is important, while the anomaly contribution in Eq. (151)
connected to the (B+L) current is irrelevant, requires an examination of the
properties of the gauge theory vacuum.  We turn to this next.

\subsection{The Gauge Theory Vacuum}

The resolution of the above issues came through a better understanding of the
vacuum structure of gauge theories (\cite{vacuum}).  The vacuum state is, by
definition, a state where all fields vanish.  For gauge fields, this needs to
be slightly extended since these fields themselves are not physical.  So,
in the case of gauge fields, the vacuum state is one where either
$A_a^\mu = 0$ or the gauge fields are a gauge transformation of
$A_a^\mu=0$.  For our purposes it suffices to examine an $SU(2)$ gauge theory,
since this example serves to exemplify what happens in a more general
case.

It proves particularly convenient (\cite{CDG}) to study the $SU(2)$ gauge
theory in a temporal gauge where $A_a^0=0~\{a=1,2,3\}$.  In this gauge the
space components of the gauge fields are time-independent $A_a^i(\vec r,t)=
A_a^i(\vec r)$.  Even so, there is still some residual gauge freedom.
Defining a gauge matrix $A^i(\vec r)$ by contracting the gauge fields with
the Pauli matrices, $A^i(\vec r) = \frac{\tau_a}{2} A_a^i(\vec r)$, in the
$A_a^0=0$ gauge one is left with the freedom to perform the following gauge
transformations
\begin{equation}
A^i(\vec r)\to\Omega(\vec r)A^i(\vec r)\Omega(\vec r)^{-1} + \frac{i}{g}
\Omega(\vec r)\nabla^i\Omega(\vec r)^{-1}~,
\end{equation}
where $g$ is the gauge coupling for the $SU(2)$ theory.  In view of the above,
one concludes that in the $A_a^0=0$ gauge, pure gauge fields corresponding
to the vacuum configuration are simply the set
$\{0,~\frac{i}{g}\Omega(\vec r)\nabla^i\Omega(\vec r)^{-1}\}$.

The behavior of $\Omega(\vec r)$ as $\vec r\to\infty$ distinguishes classes of
pure gauge fields.  In particular, the requirement that (\cite{CDG})
\begin{equation}
\Omega(\vec r)\stackrel{\vec r\to\infty}{\longrightarrow} 1,
\end{equation}
provides a map of physical space $[S_3]$ onto the group space $[SU(2)\sim S_3]$.
This $S_3\to S_3$ map splits the matrices $\Omega(\vec r)$ into different
homotopy classes $\{\Omega_n(\vec r)\}$, characterized by an integer
$n$---the winding number---specifying how $\Omega(\vec r)$ goes to unity at
spatial infinity:
\begin{equation}
\Omega_n(\vec r)\stackrel{\vec r\to\infty}{\longrightarrow} e^{2\pi in}~.
\end{equation}
Thus the set of pure gauge fields is $\{0,~A_n^i(\vec r)\}$, where
\begin{equation}
A_n^i(\vec r) = \frac{i}{g}\Omega_n(\vec r)\nabla^i\Omega_n(\vec r)^{-1}~.
\end{equation}

The winding number $n$ is just the Jacobian of the $S_3\to S_3$ transformation (\cite{Crewther}) and one can show that
\begin{equation}
n = \frac{ig^3}{24\pi^2}\int d^3r~{\rm Tr}~\epsilon_{ijk}A_n^i(\vec r)
A_n^j(\vec r)A_n^k(\vec r)~.
\end{equation}
Furthermore, one can construct the transformation matrix $\Omega_n(\vec r)$ with
winding number $n$ by compounding $n$-times the transformation matrix of unit
winding
\begin{equation}
\Omega_n(\vec r) = [\Omega_1(\vec r)]^n~.
\end{equation}
A representative $n=1$ matrix, giving rise to a, so called, {\bf large gauge
transformation} is given by
\begin{equation}
\Omega_1(\vec r) = \frac{\vec r^2-\lambda^2}{\vec r^2+\lambda^2} +
\frac{2i\lambda\vec\tau\cdot\vec r}{\vec r^2 + \lambda^2}~,
\end{equation}
with $\lambda$ an arbitrary scale parameter.

Using the above properties, it is clear that the $n$-vacuum state---corresponding
to the pure gauge field configuration $A_n^i(\vec r)$---is not fully gauge
invariant.  Indeed, a large gauge transformation can change the gauge field
$A_n^i(\vec r)$ into that of $A^i_{n+1}(\vec r)$
\begin{equation}
A^i_{n+1}(\vec r) = \Omega_1(\vec r)A_n^i(\vec r) \Omega_1^{-1}(\vec r) +
\frac{i}{g}\Omega_1(\vec r)\nabla^i\Omega_1^{-1}(\vec r)~,
\end{equation}
or
\begin{equation}
\Omega_1|n\rangle = |n+1\rangle~.
\end{equation}
The correct vacuum state for a gauge theory must be gauge invariant.  As such
it must be a linear superposition of these $n$-vacuum states.  This is the,
so-called, $\theta$-vacuum (\cite{vacuum})
\begin{equation}
|\theta\rangle = \sum_n e^{-in\theta}|n\rangle~.
\end{equation}
Clearly, since
\begin{equation}
\Omega_1|\theta\rangle = \sum_n e^{-in\theta}\Omega_1|n\rangle =
\sum_n e^{-in\theta}|n+1\rangle = e^{i\theta}|\theta\rangle~,
\end{equation}
the $|\theta\rangle$ vacuum is gauge invariant.

Using the $\theta$-vacuum as the correct vacuum state for gauge theories, it is
clear that the vacuum functional for these theories splits into distinct
sectors (\cite{CDG}).  If $|\theta\rangle_\pm$ are the $\theta$-vacuum states at
$t=\pm\infty$, then the vacuum functional for a gauge theory takes the form
\begin{eqnarray}
_+\langle\theta|\theta\rangle_- &=& \sum_{n,m}
e^{im\theta}e^{-in\theta}~_+\langle m|n\rangle_- \nonumber \\
&=&\sum_\nu e^{i\nu\theta}\left[\sum_n~_+\langle n+\nu|n\rangle_-\right]~.
\end{eqnarray}
That is, the vacuum functional sums over vacuum to vacuum amplitudes in which
the winding number at $t=\pm\infty$ differ by $\nu$, weighing each by a
factor $e^{i\nu\theta}$.  We anticipate here that the superposition of
amplitudes with {\bf different} phases $e^{i\nu\theta}$ will lead to
CP-violating effects.  Recalling that the vacuum functional is given by a path
integral over gauge field configurations, each weighted by the classical action,
one arrives at the formula
\begin{equation}
_+\langle\theta|\theta\rangle_- = \int_{\rm Paths}
\delta A_\mu e^{iS[A]} = \sum_\nu e^{i\nu\theta}
\left[\sum_\nu~_+\langle n+\nu|n\rangle_-\right]~.
\end{equation}

Although the formula for $_+\langle\theta|\theta\rangle_-$ above was derived
in the $A_a^0$ gauge, the parameter $\nu$ entering in this formula has
actually a gauge invariant meaning.  One finds (\cite{vacuum})
\begin{equation}
\nu = n_+ - n_- = \frac{g^2}{32\pi^2} \int d^4xG_a^{\mu\nu}
\tilde G_{a\mu\nu}~.
\end{equation}
To prove this result requires using Bardeen's identity (\cite{Bardeen}) which
expresses the product of $G\tilde G$ as a total derivative:
\begin{equation}
G_a^{\mu\nu}\tilde G_{a\mu\nu} = \partial_\mu K^\mu~,
\end{equation}
where the ``current" $K^\mu$ is given by
\begin{equation}
K^\mu = \epsilon^{\mu\alpha\beta\gamma}A_{a\alpha}
\left[G_{a\beta\gamma} - \frac{g}{3} \epsilon_{abc}A_{b\beta}
A_{c\gamma}\right]~.
\end{equation}
For pure gauge fields $[G_{a\beta\gamma}=0]$ and in the
$A_a^0=0$ gauge this curent has only a temporal component:
\begin{equation}
K^i=0;~K^0 = -\frac{g}{3}\epsilon_{ijk}\epsilon_{abc}
A_a^iA_b^jA_c^k = \frac{4}{3} ig\epsilon_{ijk}~{\rm Tr}~
A^iA^jA^k~.
\end{equation}
Using these relations, in this gauge one can write the winding numbers
$n_\pm$ as
\begin{equation}
n_\pm = \frac{ig^3}{24\pi^2} \int d^3r\epsilon_{ijk}~{\rm Tr}~
A^iA^jA^k = \frac{g^2}{32\pi^2} \int d^3rK^0\left|_{t=\pm\infty}\right.~.
\end{equation}
The above formula allows one to express the winding number difference
$\nu = n_+-n_-$ as
\begin{equation}
\nu = n_+-n_- = \frac{g^3}{32\pi^2}\int d^3rK^0|^{t=+\infty}_{t=-\infty}
= \frac{g^2}{32\pi^2}\int d\sigma_\mu K^\mu~.
\end{equation}
Whence, Eq. (169) follows by using Gauss's theorem and Bardeen's identity.

Having identified $\nu$ as an integral over $G\tilde G$, one can rewrite the
formula for the vacuum functional in terms of an effective action.  Defining
\begin{equation}
S_{\rm eff}[A] = S[A] + \theta\frac{g^2}{32\pi^2} \int d^4x
G_a^{\mu\nu}\tilde G_{a\mu\nu}~,
\end{equation}
one sees that
\begin{equation}
_+\langle\theta|\theta\rangle_- = \sum_\nu \int_{\rm Paths}
\delta A_\mu e^{iS_{\rm eff}[A]}\delta\left[\nu - \frac{g^2}{32\pi^2}
\int d^4xG_a^{\mu\nu}\tilde G_{a\mu\nu}\right]~.
\end{equation}
The more complicated structure of the gauge theory vacuum [$\theta$-vacuum]
effectively adds an additional term to the gauge theory Lagrangian:
\begin{equation}
{\cal{L}}_{\rm eff} = {\cal{L}}_{\rm gauge~theory} + \theta
\frac{g^2}{32\pi^2} G_a^{\mu\nu}\tilde G_{a\mu\nu}~.
\end{equation}

Perturbation theory is connected to the $\nu=0$ sector, since
$\int d^4xG\tilde G=0$.  Effects of non-zero winding number differences
$(\nu\not= 0)$ involve {\bf non-perturbative} contributions.  These are
naturally selected by the connection of the pseudoscalar density
$G\tilde G$ with the divergence of chiral currents, through the chiral
anomaly (\cite{ABJ}).

Let me examine this first for QCD.  Assuming there are $n_f$ flavors whose mass
can be neglected $(m_f=0)$, the axial current in QCD
\begin{equation}
J_5^\mu = \frac{1}{2} \sum^{n_f}_{i=1} \bar q_i\gamma^\mu\gamma_5 q_i
\end{equation}
is still not conserved as a result of the chiral anomaly.  One has
\begin{equation}
\partial_\mu J_5^\mu = n_f \frac{g_3^2}{32\pi^2}G_a^{\mu\nu}
\tilde G_{a\mu\nu}~.
\end{equation}
In view of the above, chirality changes $\Delta Q_5$, are simply related 
to $\nu$:
\begin{equation}
\Delta Q_5 = \int d^4x\partial_\mu J_5^\mu = n_f
\frac{g_3^2}{32\pi^2} \int d^4xG_a^{\mu\nu}\tilde G_{a\mu\nu} =
n_f\nu~.
\end{equation}
Clearly, if $\nu \not= 0$ sectors are important in QCD, then the above
changes are important and the corresponding $U(1)_A$ symmetry is {\bf never}
a symmetry of the theory.  This then is the physical explanation why
(in the relevant $n_f=2$ case) the $\eta$ does not have the properties of a
Goldstone boson.

't Hooft (\cite{'tHooft2}), by using semiclassical methods, provided an estimate
of the likelyhood of the occurence of
 processes involving $\nu\not= 0$ transitions.  Basically,
he viewed the transition from an $n$-vacuum at $t=-\infty$ to an
$(n+\nu)$-vacuum at $t=+\infty$ as a tunneling process and estimated the
tunneling probability by WKB methods.  't Hooft's result (\cite{'tHooft2})
\begin{equation}
A[\nu] \sim e^{-S_E[\nu]}
\end{equation}
uses as the WKB factor in the exponent the minimal Euclidean action for the
gauge theory.  Such a minimal action obtains if the gauge field configurations
are those provided by instantons (\cite{Belavin et al.}).  These are self-dual
solutions of the field equations in Euclidean space $[G_a^{\mu\nu}=
\tilde G_a^{\mu\nu}]$ and their action is simply related to $\nu$.  For these
solutions
\begin{equation}
S_E[\nu] = \frac{1}{4} \int d^4x_EG_a^{\mu\nu}G_a^{\mu\nu} = \frac{1}{4}
\int d^4x_EG_a^{\mu\nu}\tilde G_a^{\mu\nu} =
\frac{8\pi^2}{g^2_3}\nu~.
\end{equation}
What 't Hooft showed in his careful calculation (\cite{'tHooft2}) is that the
coupling constant that enters in $S_E[\nu]$ is actually a running coupling,
with its scale set by the scale of the instanton solution involved.  Further,
to evaluate the amplitude in question one must integrate over all such scales.
Thus, schematically, 't Hooft's result is
\begin{equation}
A[\nu] \sim \int d\rho~\exp\left[-\frac{2\pi\nu}{\alpha_3(\rho^{-1})}\right]~.
\end{equation}
In QCD, since the gauge coupling squared $\alpha_3(\rho^{-1})$ grows for large
distances, there is no particular suppression due to the tunneling factor
for large size instantons.  Because of this,
although one cannot really calculate $A[\nu]$, one expects that
\begin{equation}
A[\nu\not= 0] \sim A[0]~.
\end{equation}
Thus, as advertized, $U(1)_A$ is not really a symmetry of QCD.

Much of the above discussion applies to the electroweak theory.  However, as we
shall see, there is a crucial difference.  Since the electroweak theory is
based on the group $SU(2)\times U(1)$, because of the $SU(2)$ factor there is
also here a non-trivial vacuum structure.  The $W\tilde W$ density connected
to the index difference in this case is directly related to the divergence of
the B+L current.  Focusing on this contribution, one has
\begin{equation}
\partial_\mu J^\mu_{\rm B+L} = -\frac{g^2_2}{16\pi^2} N_gW_i^{\mu\nu}
\tilde W_{i\mu\nu}~.
\end{equation}
Hence, the change in (B+L) in the electroweak theory is also simply connected
to the (weak) index $\nu$ (\cite{'tHooft})
\begin{equation}
\Delta(B+L) = \int d^4x\partial_\mu J^\mu_{\rm B+L} = -\frac{g^2_2}{16\pi^2}
N_g \int d^4x W_i^{\mu\nu}\tilde W_{i\mu\nu} =
-2N_g\nu~.
\end{equation}

I note that for three generations $[N_g = 3]$ the minimal violation of the
(B+L)-current is $|\Delta({\rm B+L})|=6$.  So, even though baryon number is
violated in the Standard Model the process $p\to e^+\pi^0$, which involves
$\Delta({\rm B+L}) = 2$, is still forbidden!  More importantly, however, the
amplitude for (B+L)-violation itself is totally negligible.  This amplitude, at
least semiclassically, will again be given by a result similar to what was
obtained in QCD (except with $\alpha_3\to\alpha_2$).  However, because the 
electroweak symmetry is broken, the integration over instanton sizes
cuts off at sizes of order $1/v$ (or momentum scales of order $M_Z$).  Hence,
one estimates (\cite{'tHooft})\footnote{Here we use $\alpha_2(M_Z) =
\frac{\alpha(M_Z)}{\sin^2\theta_W}\sim \frac{1}{30}$.}
\begin{equation}
A[\nu]_{\rm (B+L)-violation} \sim \exp\left[-\frac{2\pi\nu}
{\alpha_2(M_Z)}\right] \sim 10^{-80\nu}~.
\end{equation}

I want to remark that, although the above result is negligibly small, in
the early Universe (B+L)-violation in the electroweak theory can be important.
This was first observed by Kuzmin, Rubakov, and Shaposhnikov (\cite{KRS}), who
pointed out that in a thermal bath the semiclassical estimate of
't Hooft ceases to be accurate.  Effectively, in these circumstances, the
gauge configurations associated with (B+L)-violating processes are not governed
by a tunnelling factor, but by a Boltzman factor.  As one nears the electroweak
phase transitions, furthermore, this Boltzman factor tends to unity and the
(B+L)-violating processes proceed essentially unsuppressed.

\subsection{The Strong CP Problem}

The $\theta$-vacuum of QCD is a new source of CP-violation,\footnote{One can
show that the equivalent $\theta$-parameter in the electroweak theory can be
rotated away as a result of the chiral nature of these interactions (\cite{Rubakov}).}
as a result of the effective interaction
\begin{equation}
{\cal{L}}_{\rm CP-violation} = \theta\frac{\alpha_3}{8\pi}
G_a^{\mu\nu}\tilde G_{a\mu\nu}~,
\end{equation}
which reflects the presence of the vacuum angle.  It turns out, in fact, that
the situation is a little bit more complicated, because of the electroweak
interactions.  Recall that the quark mass matrices arising as a result of the
spontaneous breakdown of $SU(2)\times U(1)$ are, in general, neither
Hermitian nor diagonal
\begin{equation}
{\cal{L}}_{\rm mass} = -\bar q_{{\rm L}i} M_{ij}q_{{\rm R}j} -
\bar q_{{\rm R}i}(M^{\dagger})_{ij}q_{{L}j}~.
\end{equation}
These matrices can, however, be diagonalized by performing appropriate unitary
transformations on the quark fields
\begin{equation}
q_{\rm R} \to q^\prime_{\rm R} = U_{\rm R}q_{\rm R}~;~~
q_{\rm L}\to q^\prime_{\rm L} = U_{\rm L}q_{\rm L}~.
\end{equation}
It is easy to check that part of the above transformations involves a
$U(1)_A$ transformation.  In fact, the $U(1)_A$ piece of these transformations
is just
\begin{eqnarray}
q_{\rm R}\to q^\prime_{\rm R} &=& \exp\left[\frac{i}{2n_f}~{\rm Arg~det}~M\right]
q_{\rm R}\equiv \exp\left[\frac{i}{2}\alpha\right] q_{\rm R}\nonumber \\
q_{\rm L}\to q^\prime_{\rm L} &=& \exp\left[-\frac{i}{2n_f}~{\rm Arg~det}~M\right]
q_{\rm L} \equiv \exp\left[-\frac{i}{2} \alpha\right] q_{\rm L}~.
\end{eqnarray}
It turns out that such $U(1)_A$ transformations engender a change in the vacuum
angle (\cite{JR}).  Thus they effectively add a contribution to Eq. (188), beyond
that of the QCD angle $\theta$.

To prove this contention (\cite{JR}), one has to examine carefully what is the
result of a chiral $U(1)_A$ transformation.  Although the current $J_5^\mu$
connected to $U(1)_A$ has an anomaly, it is always possible to construct a
conserved current by using the current $K^\mu$ which enters in Bardeen's
identity (\cite{Bardeen}).  Recalling Eqs. (170) and (179), it is obvious that
the desired conserved chiral current $\tilde J_5^\mu$ is
\begin{equation}
\tilde J_5^\mu = J_5^\mu - \frac{n_f\alpha_3}{4\pi} K^\mu~.
\end{equation}
The charge which generates chiral transformations, $\tilde Q_5$, needs to be
time-independent.  By necessity, it must therefore be related to
$\tilde J_5^\mu$---the conserved current:
\begin{equation}
\tilde Q_5 = \int d^3x\tilde J_5^0~.
\end{equation}
Although $\tilde Q_5$ is time-independent, this charge is not invariant under
large gauge transformations, since $K^\mu$ is itself not a gauge-invariant
current like $J_5^\mu$.  One finds
\begin{equation}
\Omega_1\tilde Q_5\Omega_1 = \Omega_1\left[Q_5 - \frac{n_f\alpha_3}{4\pi}
\int d^3xK^0\right] \Omega_1 = \tilde Q_5 + n_f~.
\end{equation}

Consider the action of a large gauge transformation $\Omega_1$ on a chirally
rotated $\theta$-vacuum state $e^{i\alpha\tilde Q_5}|\theta\rangle$.  One has
\begin{eqnarray}
\Omega_1\left[e^{i\alpha\tilde Q_5}|\theta\rangle\right] &=&
\Omega_1 e^{i\alpha\tilde Q_5}\Omega_1^{-1}\Omega_1|\theta\rangle \nonumber \\
&=& e^{i(\alpha n_f+\theta)}\left[e^{i\alpha\tilde Q_5}|\theta\rangle\right]~.
\end{eqnarray}
It follows from the above, immediately, that a chiral $U(1)_A$ rotation indeed
shifts the vacuum angle (\cite{JR}):
\begin{equation}
e^{i\alpha\tilde Q_5}|\theta\rangle = |\theta + \alpha n_f\rangle~.
\end{equation}
For the electroweak theory, the chiral rotation one needs to perform to
diagonalize the quark mass matrices has a parameter $\alpha = \frac{1}{n_f}
{\rm det}~M$.  Whence, it follows that the effective CP-violating Lagrangian
term arising from the structure of the gauge theory vacuum is
\begin{equation}
{\cal{L}}^{\rm eff}_{\rm CP-violation} = \bar\theta\frac{\alpha_3}{8\pi}
G_a^{\mu\nu}\tilde G_{a\mu\nu}~,
\end{equation}
where
\begin{equation}
\bar\theta = \theta + {\rm Arg~det}~M~.
\end{equation}
The effective CP-violating parameter $\bar\theta$ is the sum of a QCD
contribution---the vacuum angle $\theta$---and an electroweak piece--Arg det
$M$---related to the phase structure of the quark mass matrix.

The interaction (197) is C even, and T and P odd.  Thus it violates CP also.
It turns out, as we shall see below, that unless $\bar\theta$ is very small
$[\bar\theta \leq 10^{-10}]$ this interaction produces an electric dipole
moment for the neutron which is beyond the present experimental bound for this
quantity.  It is difficult to understand why a parameter like $\bar\theta$,
which is a sum of two very different contributions, should be so small.  This
conundrum is known as the strong CP problem.

Before discussing the strong CP problem further, let me first indicate how
to calculate the contribution of the effective Lagrangian (197) to the
electric dipole moment of the neutron.  This is most easily done by transforming
the $\bar\theta$ interaction from an interaction involving gluons to one
involving quarks.  For simplicity, let me concentrate on the two-flavor case
$(n_f=2)$ and take, again for simplicity, $m_u=m_d=m_q$.  In this case, it is
easy to see that the chiral $U(1)_A$ transformation
\begin{equation}
\left( \begin{array}{c}
u \\ d
\end{array} \right) \to
\exp\left[i\frac{\bar\theta\gamma_5}{4}\right]
\left( \begin{array}{c}
u \\ d
\end{array} \right)
\end{equation}
will get rid of the $\bar\theta G\tilde G$ term.  However, the above
transformation will, at the same time, generate a CP-violating
$\gamma_5$-dependent mass term for the $u$ and $d$ quarks:
\begin{equation}
{\cal{L}}_{\rm CP-violation}^{\rm eff} = i\bar\theta m_q
\left[\bar u\frac{\gamma_5}{2} u + \bar d \frac{\gamma_5}{2} d\right]~.
\end{equation}
One can use the above effective Lagrangian directly to calculate the neutron
electric dipole moment.  One has, in general
\begin{equation}
d_n\bar n\sigma_{\mu\nu} k^\nu\gamma_5n = 
\langle n|T(J^{\rm em}_\mu i\int d^4x{\cal{L}}^{\rm eff}_{\rm CP-violation})
|n\rangle~.
\end{equation}
To arrive at a result for $d_n$ one inserts a complete set of states
$|X\rangle$ in the matrix element above and tries to estimate which set of
states $|X\rangle$ dominates.  In the literature there are two calculations
along these lines.  Baluni (\cite{Baluni}) uses for $|X\rangle$ the odd parity
$|N^-_{1/2}\rangle$ states which are coupled to the neutron by
${\cal{L}}^{\rm eff}_{\rm CP-violation}$.  Crewther {\it et al.} (\cite{CDVW}),
instead, do a soft pion calculation (effectively $|X\rangle\sim |N\pi_{\rm soft}\rangle$).  The result of these calculations are rather similar and lead
to an expression for $d_n$ whose form could have been guessed.  Namely
\begin{equation}
d_n\sim \frac{e}{M_n} \left(\frac{m_q}{M_n}\right) \bar\theta\sim
\left\{ \begin{array}{l}
2.7\times 10^{-16}~\bar\theta~{\rm ecm}~(\rm{Baluni~1979}) \\
5.2\times 10^{-16}~\bar\theta~{\rm ecm}~(\rm{Crewther~et~al~1979})
\end{array} \right.
\end{equation}
The present bound on $d_n$ (\cite{PDG}) is, at 95\% C.L.,
\begin{equation}
d_n < 1.1\times 10^{-25}~{\rm ecm}~.
\end{equation}
Whence, to avoid contradictions with experiment, the parameter $\bar\theta$
must be less than $2\times 10^{-10}$.  Why this should be so is a mystery.
This is the strong CP problem.

\subsection{The Chiral Solution to the Strong CP Problem}

About twenty years ago, Helen Quinn and I (\cite{PQ}) suggested a possible
dynamical solution to the strong CP problem.  If our mechanism holds in
nature then $\bar\theta$ actually vanishes, and there is no need to explain a
small numbr like $10^{-10}$ cropping up in the theory.\footnote{Even 
incorporating a $U(1)_{\rm PQ}$ symmetry into the theory
it turns out that 
CP violating effects in the electroweak interactions do not allow
$\bar\theta$ to totally  
vanish.  However, the effective $\bar\theta$ induced back
through weak CP-violation is tiny $(\bar\theta\sim 10^{-15})$ (\cite{Georgi})
and well within the bound provided by the neutron electric dipole moment.}
To ``solve" the strong CP problem, Quinn and I postulated that the
Lagrangian of the Standard Model was invariant under an additional global
$U(1)$ chiral symmetry---$U(1)_{\rm PQ}$.  This required imposing certain
constraints on the Higgs sector of the theory, but otherwise appeared
perfectly possible.  Because the $U(1)_{\rm PQ}$ symmetry is a chiral symmetry,
if this symmetry were exact, it is trivial to see that the $\bar\theta G\tilde G$
term can be eliminated, since the chiral rotation $\exp\left[-i\frac{\theta}{n_f}
\tilde Q_5^{\rm PQ}\right]$ gives
\begin{equation}
\exp\left[-i\frac{\bar\theta}{n_f}\tilde Q_5^{\rm PQ}\right]|\bar\theta\rangle
= |0\rangle~.
\end{equation}
That is, by a $U(1)_{\rm PQ}$ transformation the effective vacuum angle
$\bar\theta$ is set to zero and this parameter 
is no longer present in the theory.  Phyically, however, if
$U(1)_{\rm PQ}$ is an extra global symmetry of the Standard Model, it is not
possible for this symmetry to remain unbroken.  What Quinn and I showed (\cite{PQ})
was that, even if $U(1)_{\rm PQ}$ is spontaneously broken, one still is able
to eliminate the $\bar\theta G\tilde G$ term.

To see this, it is useful to focus on the associated Nambu-Goldstone boson
resulting from the spontaneous breakdown of the $U(1)_{\rm PQ}$ symmetry.
This excitation is the axion, first discussed by Weinberg and Wilczek (\cite{WeW}) in connection with the $U(1)_{\rm PQ}$ symmetry.  It
turns out that the axion is not quite massless, so it is really a
pseudo-Goldstone boson (\cite{pseudo}).  This is a consequence of the
$U(1)_{\rm PQ}$ symmetry having an anomaly due to QCD interactions.
One finds (\cite{WeW}) that the axion mass is of order
\begin{equation}
m_a \sim \frac{\Lambda^2_{\rm QCD}}{f}~,
\end{equation}
where $\Lambda_{\rm QCD}$ typifies the scale of the QCD interactions, while
$f$ is the scale of the $U(1)_{\rm PQ}$ breakdown.  If $f \gg \Lambda_{\rm QCD}$, then axions turn out to be very much lighter than ordinary hadrons.

If we denote the axion field by $a(x)$, it turns out that imposing a
$U(1)_{\rm PQ}$ symmetry on the standard model effectively serves to
replace the CP-violating $\bar\theta$ parameter by the dynamical
CP-conserving axion field:
\begin{equation}
\bar\theta\to \frac{a(x)}{f}~.
\end{equation}
To understand why this is so, recall that since the axion is the
Nambu-Goldstone boson of the broken $U(1)_{\rm PQ}$ symmetry, this field
translates under a $U(1)_{\rm PQ}$ transformation:
\begin{equation}
a(x)\stackrel{U(1)_{\rm PQ}}{\longrightarrow} a(x) + \alpha f~,
\end{equation}
where $\alpha$ is the parameter associated with the $U(1)_{\rm PQ}$
transformation.  Because of Eq. (207), the axion field can only enter in
the Lagrangian of the theory through derivative terms.  Even though the 
detailed axion interactions are somewhat model-dependent, this property
allows one to understand how to augment the Lagrangian of the Standard
Model so that it becomes $U(1)_{\rm PQ}$ invariant. 

 Focussing only on the
possible additional contributions due to the inclusion of the axion field,
one is lead to the following effective Lagrangian for the theory
\begin{eqnarray}
{\cal{L}}^{\rm eff}_{\rm SM} = {\cal{L}}_{\rm SM} &+& \bar\theta
\frac{\alpha_3}{8\pi} G_a^{\mu\nu}\tilde G_{a\mu\nu} - \frac{1}{2}
\partial_\mu a\partial^\mu a \nonumber \\
&+& {\cal{L}}^{\rm int}_{\rm axion} \left[\frac{\partial_\mu a}{f}~;
\psi\right] + \frac{a}{f} \xi\frac{\alpha_3}{8\pi}
G_a^{\mu\nu}\tilde G_{a\mu\nu}~.
\end{eqnarray}
The third term above is the kinetic energy term for the axion field, while
the fourth term in Eq. (208) schematically indicates the kind of
interactions the axion field can participate in with the other fields
$[\psi]$ in the theory.  The last term above, as can be noticed, does not
involve a derivative of the axion field, thereby violating the usual
expectations for Nambu-Goldstone fields.  The reason why this term is
included, however, is clear.  The $U(1)_{\rm PQ}$ symmetry is
anomalous\footnote{Here $\xi$ is a model-independent number of $O(1)$ (see, for example, \cite{RDP}).}
\begin{equation}
\partial_\mu J^\mu_{\rm PQ} = \xi\frac{\alpha_3}{8\pi}
G_a^{\mu\nu}\tilde G_{a\mu\nu}~.
\end{equation}
This anomaly must be reflected in the effective Lagrangian (208) when one
performs a chiral $U(1)_{\rm PQ}$ transformation.  This is guaranteed by
having the last term in Eq. (208), since it precisely reproduces the anomaly
when the axion field undergoes the $U(1)_{\rm PQ}$ transformation (207).

The last term in Eq. (208), whose origin is intimately connected to the
chiral anomaly, because it contains the axion field directly (and not its
derivative) provides a potential for the axion field.  As a result, it is
not true anymore that all values of the vacuum expectation value (VEV) of
$a(x)$ are allowed.\footnote{This would be true if ${\cal{L}}^{\rm eff}_{\rm SM}$ only contained interactions involving $\partial_\mu a$,
since these cannot fix a value for the VEV of $a,\langle a\rangle$.}
The minimum of $V_{\rm eff}$ in the vacuum is simply
\begin{equation}
\left\langle\frac{\partial V_{\rm eff}}{\partial_a}\right\rangle = -
\frac{\xi}{f}\frac{\alpha_3}{8\pi}
\langle G_a^{\mu\nu}\tilde G_{a\mu\nu}\rangle\left|_{\langle a\rangle
\not= 0}\right.~.
\end{equation}
What Quinn and I showed (\cite{PQ}), in essence, is that the periodicity of
$\langle G\tilde G\rangle$ in the effective vacuum angle
$\theta_{\rm eff}$ for the Lagrangian of Eq. (208)
\begin{equation}
\theta_{\rm eff} = \bar\theta + \frac{\xi}{f}
\langle a(x)\rangle~,
\end{equation}
requires that $\theta_{\rm eff} = 0$, or
\begin{equation}
\langle a(x)\rangle = -\frac{f}{\xi} \bar\theta~.
\end{equation}
As a result of Eq. (212), only the physical axion field
\begin{equation}
a(x)_{\rm phy} = a(x)-\langle a(x)\rangle
\end{equation}
interacts with the gluon field strengths, eliminating altogether the
$\theta G\tilde G$ term.  Thus, indeed, imposing an additional
$U(1)_{\rm PQ}$ symmetry in the Standard Model, even in the case this
symmetry is spontaneously broken, solves the strong CP problem.

As we remarked earlier, the axion is actually massive because of the
anomaly in the $U(1)_{\rm PQ}$ current.  This follows readily from the
effective Lagrangian (208).  The second derivative of the effective
potential $V_{\rm eff}$, which arose precisely because of the chiral
anomaly in the $U(1)_{\rm PQ}$ symmetry, when evaluated at its minimum
value $\langle a(x)\rangle$ gives for the axion mass squared the value
\begin{equation}
m_a^2 = \left.\left\langle\frac{\partial^2 V_{\rm eff}}{\partial a^2}\right\rangle
\right|_{\langle a\rangle} = -\frac{\xi}{f}
\frac{\alpha_3}{8\pi} \frac{\partial}{\partial a}
\left.\langle G_a^{\mu\nu}\tilde G_{a\mu\nu}\rangle\right|_{\langle a\rangle} \sim
\frac{\Lambda^2_{\rm QCD}}{f}~.
\end{equation}
Using the above results, it is clear that the effective theory incorporating
$U(1)_{\rm PQ}$ and axions no longer suffers from the strong CP problem.
All that remains as a signal of this erstwhile problem is the direct
interaction of the (massive) axion field with the gluonic pseudoscalar
density.
\begin{eqnarray}
{\cal{L}}^{\rm eff}_{\rm SM} = {\cal{L}}_{\rm SM} &+&
{\cal{L}}^{\rm int}_{\rm axion} \left[\frac{\partial_\mu a_{\rm phys}}{f}~;~
\psi\right] - \frac{1}{2} \partial_\mu a_{\rm phys}\partial^\mu
a_{\rm phys} \nonumber \\
&-& \frac{1}{2} m^2_a a^2_{\rm phys} + \frac{a_{\rm phys}}{f} \xi
\frac{\alpha_3}{8\pi} G_a^{\mu\nu}\tilde G_{a\mu\nu}~.
\end{eqnarray}

As is obvious from the above equation, the physics of axions depends on the
scale of $U(1)_{\rm PQ}$ breaking $f$.  In the original model Helen Quinn
and I put forth (\cite{PQ}), we associated $f$ quite naturally with the scale
of electroweak symmetry breaking $v = (\sqrt{2}~G_F)^{-1/2}$.  To impose
the $U(1)_{\rm PQ}$ symmetry on the Standard Model we had to have two
distinct Higgs doublets, $\Phi_1$ and $\Phi_2$, with different $U(1)_{\rm PQ}$
charges.  The axion field then turns out to be the common phase field
of $\Phi_1$ and $\Phi_2$ which is orthogonal to the weak 
hypercharge (\cite{RDP}).  Isolating just this contribution in $\Phi_1$ and
$\Phi_2$, one has
\begin{equation}
\Phi_1 = \frac{v_1}{\sqrt{2}} 
\exp\left[ix\frac{a}{f}\right] 
\left( \begin{array}{c}
1 \\ 0
\end{array} \right)~; ~~
\Phi_2 = \frac{v_2}{\sqrt{2}} \exp\left[i\frac{a}{xf}\right]
\left( \begin{array}{c}
0 \\ 1
\end{array} \right)~.
\end{equation}
Here $x=v_2/v_1$, is the ratio of the two Higgs VEV's and the
$U(1)_{\rm PQ}$ symmetry breaking scale $f$ is given by
\begin{equation}
f=\sqrt{v_1^2+v_2^2} = (\sqrt{2}~G_F)^{-1/2} \simeq
250~{\rm GeV}~.
\end{equation}

The $\Phi_1$ field has weak hypercharge of $-1/2$, while the $\Phi_2$ field
has weak hypercharge of +1/2.  Hence, in the Yukawa interactions $\Phi_1$
couples the $u_{{\rm R}j}$ fields to the left-handed quark doublets,
while $\Phi_2$ couples $d_{{\rm R}j}$ to these same fields
\begin{equation}
{\cal{L}}_{\rm Yukawa} = -\Gamma^u_{ij}(\bar u,\bar d)_{{\rm L}i}
\Phi_1u_{{\rm R}j} - \Gamma^d_{ij}(\bar u,\bar d)_{{\rm L}i}
\Phi_2d_{{\rm R}j} + \hbox{h.c.}
\end{equation}
In view of Eq. (216), it is clear that the above interaction is
$U(1)_{\rm PQ}$ invariant.  The shift of the axion field by $\alpha f$
[cf Eq. (207)] under a $U(1)_{\rm PQ}$ transformation is compensated by
an appropriate rotation of the right-handed quark fields. Specifically, under a
$U(1)_{\rm PQ}$ transformation one has
\begin{eqnarray}
a_{\rm phys} \stackrel{PQ}{\longrightarrow} a_{\rm phys} +\alpha f 
\nonumber \\
u_{{\rm R}j} \stackrel{PQ}{\longrightarrow} \exp \left[-i\alpha x\right]u_{{\rm R}j} \nonumber \\
d_{{\rm R}j}\stackrel{PQ}{\longrightarrow} \exp\left[-i\frac{\alpha}{x}\right] d_{{\rm R}j}~.
\end{eqnarray}
It is clear from the above that this $U(1)_{\rm PQ}$ transformation
encompasses also a $U(1)_A$ transformation.  As a result, one can use
$U(1)_{\rm PQ}$ to send $\bar\theta\to 0$, as advertized.

Unfortunately, weak interaction scale axions [with $f\sim 250~{\rm GeV};~
m_a \sim 100~{\rm keV}$] of the type which ensue in the model suggested
by Helen Quinn and myself, or in variations thereof, have been
ruled out experimentally.  I do not want to review all the relevant data here,
as this is done already fully elsewhere (\cite{RDP}).  An example, however,
will give a sense of the strength of this assertion.  If weak scale axions were to
exist, one expects a rather sizable branching ratio for the decay
$K^\pm \to \pi^\pm a$ (\cite{BPY})
\begin{equation}
BR(K^\pm\to \pi^\pm a)\sim 3\times 10^{-5}~.
\end{equation}
Experimentally, however, the process $K^+\to\pi^+$ ``Nothing", which would
reflect the axion decay of the $K^+$ meson, has a bound roughly three orders
of magnitude lower (\cite{KEK})
\begin{equation}
BR(K^+\to \pi^+ +~\hbox{Nothing}) < 3.8\times 10^{-8}~.
\end{equation}
One can bypass this bound by modifying the $U(1)_{\rm PQ}$ properties of the
Higgs fields involved.  However, these variant model themselves run into
other experimental troubles (\cite{RDP}).

Although weak scale axions do not exist, it is still possible that the
strong CP problem is solved because of the existence of a $U(1)_{\rm PQ}$
symmetry.  The dynamical adjustment of $\bar\theta\to 0$ works 
{\bf independently} of what is the scale, $f$, of the spontaneous symmetry
breaking of $U(1)_{\rm PQ}$.  Obviously if $f \gg (\sqrt{2}~G_F)^{-1/2}$,
the resulting axions are extremely light $(m_a\sim\Lambda^2_{\rm QCD}/f)$,
extremely weakly coupled (couplings $\sim f^{-1}$) and very long lived
$(\tau_a\sim f^5)$ and thus are essentially {\bf invisible}.  A variety of
invisible axion models have been suggested in the literature (\cite{invisible})
and they offer an interesting, if perhaps unconventional, resolution of
the strong CP problem.  Fortunately, as we shall see, these models are
actually testable.

If $f \gg (\sqrt{2}~G_F)^{-1/2}$, it is clear that the spontaneous breakdown
of $U(1)_{\rm PQ}$ must occur through a VEV of a field which is an
$SU(2)\times U(1)$ singlet.  Thus, in invisible axion models, the axion
is essentially the phase associated with an $SU(2)\times U(1)$ singlet
field $\sigma$.\footnote{The field $\sigma$ need not necessarily be an
elementary scalar field (\cite{Kim}).}  Keeping only the axion degrees of
freedom, one has
\begin{equation}
\sigma = \frac{f}{\sqrt{2}} e^{ia/f}~.
\end{equation}
It turns out that astrophysics and cosmology give important constraints
on the $U(1)_{\rm PQ}$ breaking scale $f$, or equivalently the axion
mass (\cite{RDP})
\begin{equation}
m_a\simeq 6\left[\frac{10^6~{\rm GeV}}{f}\right]~{\rm eV}~.
\end{equation}
These constraints restrict the available parameter space for invisible
axion models and suggest ways in which these excitations, if they exist,
could be detected.  Let me briefly discuss these matters.

The astrophysical bounds on axions arise because, if $f$ is not large
enough, axion emission removes energy from stars, altering their evolution.
These bounds are reviewed in great details in a recent monograph by 
Raffelt (\cite{Raffelt}).  Although these bounds are somewhat dependent on the
type of invisible axion model one is considering, typically invisible axions
avoid all astrophysical constraints if
\begin{equation}
f\geq 5\times 10^9~{\rm GeV}~; ~~
m_a\leq 10^{-3}~{\rm eV}~.
\end{equation}
Cosmology, on the other hand, provides an upper bound on $f$ (\cite{cosmologybound}).  At the $U(1)_{\rm PQ}$ phase transition in the
early Universe, at temperatures $T\sim f$, the effects of the QCD anomaly
are not yet felt and the axion vacuum expectation value $\langle a\rangle$
is not alligned dynamically to cancel the $\bar\theta$ term.  This
cancellation only occurs as the Universe cools towards temperatures $T$
of order $T\sim \Lambda_{\rm QCD}$.  The axion VEV $\langle a\rangle$,
as the temperature decreases, is driven to the correct minimum in an
oscillatory fashion.  These coherent, zero momentum, axion oscillations
contribute to the Universe's energy density.  If $f$ is too large, in fact,
the energy density due to axions can overclose the Universe.  Demanding
that this not happen gives a bound (\cite{cosmologybound}):
\begin{equation}
f\leq 10^{12}~{\rm GeV}~; ~~
m_a\geq 6\times 10^{-6}~{\rm eV}~.
\end{equation}
This bound has some uncertainties, related to cosmology (for a discussion see, for example, \cite{Korea}), but
otherwise is not very dependent on the properties of the invisible axions
themselves.

If axions contribute substantially to the Universe's energy density, the value of $f$ (or $m_a$) will be close to the above bound.  If this is the case,
axions could be the source for the dark matter in the Universe.  Remarkably,
then, it may be actually possible, experimentally, to detect signals for
these invisible axions.  The basic idea, due to Sikivie (\cite{Sikivie}),
is to try to convert axions, trapped in the galactic halo, into photons in
a laboratory magnetic field.

If invisible axions constitute the dark matter of our galactic halo, they
would have a velocity typical of the virial velocity in the galaxy,
$v_a\sim 10^{-3}{\rm c}$.  Further, as the dominant components of the
energy density of the Universe, axions would have a typical energy density
in the halo of order
\begin{equation}
\rho^{\rm halo}_a \sim 5\times 10^{-25}~{\rm g/cm}^3 \sim
300~{\rm MeV/cm}^3~.
\end{equation}
As a result of the (electromagnetic) anomaly, axions have an interaction
with the electromagnetic field given by the effective Lagrangian (\cite{RDP})
\begin{equation}
{\cal{L}}^{\rm eff}_{a\gamma\gamma} = \frac{\alpha}{\pi f}
K_{a\gamma\gamma} a\vec E\cdot\vec B~.
\end{equation}
Here $K_{a\gamma\gamma}$ is a model dependent parameter of $O(1)$.
As a result of the above interaction, in the presence of an external
magnetic field a galactic axion can convert into a photon.

\begin{figure}
\begin{center}
\epsfig{file=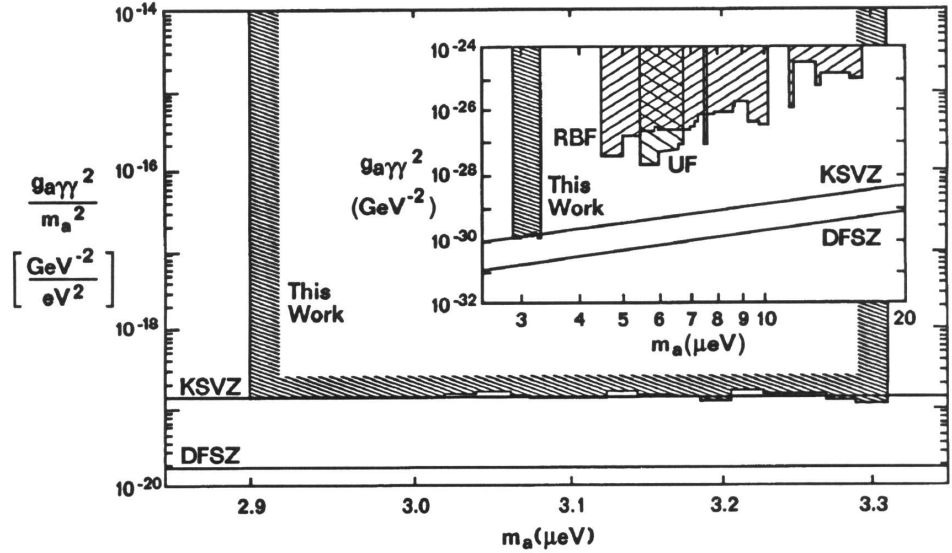,width=5in}
\end{center}
\caption[]{Result of the Livermore experiment, along with limits from some 
previous axion searches}
\end{figure}

Specifically, the electric field produced by an axion of energy
$E_a\simeq m_a$ in the presence of a magnetic field $\vec B_0$ can be
deduced from the modified wave equation
\begin{equation}
\left(\vec\nabla^2 - \frac{\partial^2}{\partial t^2}\right)
\vec E = \frac{\alpha}{\pi f} K_{a\gamma\gamma}\vec B_o
\frac{\partial^2a}{\partial t^2}~.
\end{equation}
Experimentally, the generated electromagnetic energy can be detected by
means of a resonant cavity.  When the cavity is tuned to the axion
frequency $w_a\simeq m_a$, one should get a narrow line on top of the
noise spectrum.  On resonance, the axion to photon conversion power is
given by the expression (\cite{Sikivie})
\begin{equation}
P_{\rm axion} = \frac{\rho_a}{m_a}\cdot VB_o^2\cdot
\left[\frac{\alpha}{\pi f} K_{a\gamma\gamma}\right]^2
C_{\rm overlap}\cdot Q_{\rm eff}~.
\end{equation}
In the above, the first factor gives the expected number of axions per unit
volume, the second details the magnetic energy stored in the cavity, the
third contains the coupling strength squared, $g^2_{a\gamma\gamma} =
\left[\frac{\alpha}{\pi f} K_{a\gamma\gamma}\right]^2$.  Finally,
$C_{\rm overlap}\simeq 0.7$ is an effectiveness factor for the cavity and
$Q_{\rm eff}$ is the least value between the $Q$ of the cavity itself
$[Q\sim 10^6]$ and the $Q$ due to the energy spread in the spectrum of
halo axion, $Q_a \simeq [v_a^2/c^2]^{-1}\sim 10^6$.

Halo axions produce microwave photons, since $4\times 
10^{-6}~{\rm eV}~ \equiv
1~{\rm GHz}$.  Two pilot experiments carried out in the late 80's (\cite{pilot})
had limited magnetic energy $[VB_o^2\simeq 0.5~{\rm m}^3~\hbox{(Tesla)}^2]$ and
relatively noisy amplifiers.  These experiments set limits for $g^2_{a\gamma\gamma}$ about 2 to 3 orders of magnitude above the theoretical
expectations.  Presently, there are two second generation
experiments underway, one at the Lawrence Livermore National Laboratory and
the other in Kyoto.  The Livermore experiment uses a very large
$VB_o^2\sim 12~{\rm m}^3~\hbox{(Tesla)}^2$ and low noise ``state-of-the-art"
amplifiers.  Although the signal expected at 1 GHz is tiny,
$P_{\rm axion} \sim 5\times 10^{-22}$ Watts, this experiment has already
excluded a set of invisible axion masses, at the level of strength expected
theoretically.  These recent results (\cite{Livermore}), along with some of the
older data are shown in Fig. 2.  The Kyoto experiment (\cite{Kyoto}) uses a
moderate $VB^2_o\sim 0.2~m^3~\hbox{(Tesla)}^2$.  However, it utilizes an
extremely clever technique for counting the number of photons converted
from axions---using Rydberg atoms---which makes up for the small
$VB_o^2$.  The Kyoto experiment is presently in a testing phase.  One hopes
that when both the Livermore and Kyoto experiments are completed, in 3-5
years time, they will have settled the important question of whether axions
exist or not.

\subsection{Do Real Nambu-Goldstone Bosons Exist?}

We have known for almost 40 years that when a global symmetry group $G$
breaks down spontaneously to a subgroup $H~[G\to H]$, dim $G/H$ massless
Nambu-Goldstone bosons (\cite{NG}) appear in the spectrum of the theory.
However, we have no {\bf real} physical examples still of this phenomena.
To be fair, pions are an excellent example of states which are {\bf nearly}
Nambu-Goldstone bosons.  However, although there is no question that pions
are the Nambu-Goldstone 
excitations associated with the breakdown of the
$SU(2)_V\times SU(2)_A$ approximate global symmetry of QCD to
$SU(2)_V$, pions have a small mass since the $u$ and $d$ quarks are not
exactly massless.

For a while, it was believed that it was impossible for real physical
Nambu-Goldstone bosons to exist in nature.  The argument was simple.  Because
these particles are massless, their existence seemed to be precluded by
the fact that the only long-range forces we know in nature are gravity and
electromagnetism.  However, in the early 1980's it was realized that the
existence of $m=0$ Nambu-Goldstone bosons does not pose a contradiction,
so that one can actually contemplate the interesting possibility that such
states may actually exist.

This idea came up first as a result of studying the possibility that lepton
number may be spontaneously violated.  Chikashige, Mohapatra,
and I (\cite{CMP}) dubbed the Nambu-Goldstone boson associated with the
spontaneous breakdown of lepton number a Majoron.  Soon thereafter, others  (\cite{Familon}) considered theories where one had a global
family number which could also be spontaneously broken, resulting in
other types of real Nambu-Goldstone bosons, given the name of Familons.

In this subsection I want to explain briefly why, in general,
 real Nambu-Goldstone
bosons, are not dangerous excitations to have in a theory.
After having done so, I then want to discuss briefly a specific type of
Majoron model, to illustrate some of the consequences of these kind of
models.  Succintly, the reason why Nambu-Goldstone bosons do not run afoul
with present limits on possible additional long-range forces is due to a
little theorem of Gelmini, Nussinov, and Yanagida (\cite{GNY}), which shows
that the exchange of Nambu-Goldstone bosons leads only to a long-range
tensor force.

The proof of the Gelmini-Nussinov-Yanagida theorem is very simple.  One is
interested in the potential produced by the exchange of a Nambu-Goldstone
boson between two fermions.  Recall that Nambu-Goldstone boson fields,
$\pi$, always shift under a broken symmetry transformation [cf Eq. (207)
for the axion].  Therefore, one has
\begin{equation}
\pi(x)\stackrel{\xi}{\longrightarrow} \pi(x) + v_\pi\xi~.
\end{equation}
Here $\xi$ is a parameter in $G/H$ and $v_\pi$ is a scale parameter
associated with the symmetry breakdown in question.  As a result of
Eq. (230), clearly Nambu-Goldstone fields must always be derivatively
coupled.  Hence, the most general coupling of a Nambu-Goldstone boson $\pi$
to two fermions $f_1$ and $f_2$ takes the form
\begin{equation}
{\cal{L}}^{\rm fermion}_{\rm NGB} = i\frac{\partial_\mu\pi}{v_\pi}
\bar f_1[a\gamma_\mu + b\gamma_\mu\gamma_5]f_2 + \hbox{h.c.}~,
\end{equation}
where $a$ and $b$ are numerical coefficients.  If one uses the fermion
equations of motion, one can reduce the above to a more useful form 
involving the $\pi$ field directly
\begin{equation}
{\cal{L}}^{\rm fermion}_{\rm NGB} = \frac{\pi}{v_\pi} \bar f_1
[a(m_1-m_2) + b(m_1+m_2)\gamma_5]f_2 + \hbox{h.c.}~,
\end{equation}
where $m_1$ and $m_2$ are the masses of the fermion fields $f_1$ and
$f_2$, respectively.

In calculating the potential due to $\pi$-exchange between two fermions
one needs, at each vertex, to use the interaction Lagrangian above with
$f_1=f_2$.  Obviously, for two equal fermions, the effective coupling of a
Nambu-Goldstone boson is always a {\bf pseudoscalar} coupling.  Thus
Nambu-Goldstone boson exchange cannot really generate coherent long-range
forces, since a pseudoscalar coupling in the non-relativistic limit reduces
to a $\vec\sigma\cdot\vec p$ coupling.  More precisely, the effective diagonal
coupling of a Nambu-Goldstone boson, $\pi$, to a fermion, $f$, is given by
\begin{equation}
{\cal{L}}^{\rm diag}_{\rm NGB} = ig_\pi \frac{m_f}{v_\pi} 
\bar f\gamma_5 f_\pi~,
\end{equation}
where $g_\pi$ is a, dimensionless, coupling constant.  In the non-relativistic
limit, the above reduces to
\begin{equation}
{\cal{L}}^{\rm diag}_{\rm NGB} \to g_\pi\chi^*_f
\frac{\vec\sigma\cdot\vec\nabla}{v_\pi} \chi_f\pi~,
\end{equation}
where $\chi_f$ is a Pauli spinor.  Such an interaction gives an exchange
potential between two fermions which is spin-dependent and tensorial,
with an $1/r^3$ not an $1/r$ fall off
\begin{equation}
V^{\rm eff}_{\rm NGB-exchange} = \frac{g_\pi^2/4\pi}{v_\pi^2}
\left\{\frac{\vec\sigma_1\cdot\vec\sigma_2-3(\vec\sigma_1\cdot\hat r)
(\vec\sigma_2\cdot\hat r)}{r^3} + \frac{4\pi}{3}
\delta^3(r)\vec\sigma_1\cdot\vec\sigma_2\right\}~.
\end{equation}

There have been analyses in the literature (\cite{FS}) of the size of possible
non-magnetic dipole-dipole interactions in matter, precisely of the type
one would obtain from the exchange of a real Nambu-Goldstone boson.
These bounds effectively limit how small the scale $v_\pi$ can be.
One finds no contradiction with experiment (\cite{CMP}) provided 
that
\begin{equation}
\frac{v_\pi}{g_\pi} \geq {\rm TeV}~.
\end{equation}
Thus, one can contemplate having real Nambu-Goldstone bosons of global
symmetries which are broken down at scales not much bigger than the weak
scale! If $v_\pi/g_\pi$ is much above the bound (236), clearly one expects
no measurable effects in matter.  Furthermore, if $v_\pi/g_\pi$ is
large, these Nambu-Goldstone bosons are also hard to directly produce,
since the effective coupling for producing them from a fermion $f$ scales
like $g_\pi m_f/v_\pi$.

\subsection{Majorons.}

I want to illustrate the above discussion by briefly considering the simplest
example of spontaneously broken Lepton number and its associated Majoron.
As we discussed earlier, Lepton number is a classical global symmetry of the
Standard Model.  Even at the quantum level, because the $\nu\not= 0$
amplitudes are highly suppressed, this remains an almost exact symmetry.
However, there is no reason why Lepton number should remain a symmetry of
the theory, once one considers extensions of the Standard Model.  Indeed,
the simplest extension of the Standard Model introduces right-handed
neutrino fields $\nu_{{\rm R}i}$ for each family.  Because these fields are
$SU(2)\times U(1)$ singlets, one can write a Majorana (fermion-fermion)
mass term involving these fields of the form
\begin{equation}
{\cal{L}}_{\rm mass} = -\frac{(M_{\rm R})_{ij}}{2}
\nu^T_{{\rm R}i}C\nu_{{\rm R}j} + \hbox{h.c.}~,
\end{equation}
with $C$ being the charge conjugation matrix introduced earlier
($C=1$ in the Majorana representation).  Obviously ${\cal{L}}_{\rm mass}$
does not respect Lepton number, since its two terms carry Lepton number
+2 and -2, respectively.

One can restore Lepton number as a symmetry in the above example by
introducing an appropriately transforming Higgs field.  In this case, what
one needs is a complex $SU(2)\times U(1)$ singlet field $\sigma$, which
carries Lepton number -2 (\cite{CMP}).  Clearly the interaction Lagrangian
\begin{equation}
{\cal{L}}_{\rm CMP} = -\frac{h_{ij}}{\sqrt{2}}
\left[\nu^T_{{\rm R}i} C\nu_{{\rm R}j}\sigma + \hbox{h.c.}\right]
\end{equation}
is $L$ invariant by construction.  If the dynamics of the theory forces
$\sigma$ to acquire a VEV, $\langle\sigma\rangle = \frac{1}{\sqrt{2}} V$,
then the above Lagrangian reproduces the effect of having an explicit
Majorana mass term for the right-handed neutrino fields.  In this case, one
has
\begin{equation}
(M_{\rm R})_{ij} = h_{ij}V~,
\end{equation}
and Lepton number is spontaneously broken.  Hence this
theory must also contain an explicit Nambu-Goldstone boson---the Majoron.
This is the model which I first studied with Chikashige and
Mohapatra (\cite{CMP}).

As was the case for the axion, the Majoron can also be identified here
as the phase field associated with the complex field $\sigma$.  Focusing
only on the Majoron, $\chi$, degrees of freedom, one can write
\begin{equation}
\sigma\simeq\frac{V}{\sqrt{2}} e^{i\chi/V}~.
\end{equation}
If the interaction (238) was the only interaction that the $\nu_{{\rm R}i}$
fields had, then clearly $\chi$ would couple only to these fields.  However,
once one introduces right-handed neutrino fields, one cannot avoid coupling
$\nu_{{\rm R}i}$ to the usual leptonic doublet fields $(\nu,e)_{{\rm L}i}$
via the ordinary Higgs doublet field.  As a result of these couplings, the
Majoron field $\chi$ also ends up by having a (small) interaction with
the left-handed neutrino fields.  However, if the right-handed Majorana mass
$M_{\rm R}$ (or, equivalently, the VEV of the $\sigma$-field $V$) is large,
the Majoron still predominantly couples to the right-handed neutrinos.

Let us see how this goes in detail.  As a result of the spontaneous breaking
of both Lepton number and $SU(2)\times U(1)$, the neutrino fields have both
a Dirac (fermion-antifermion) and a Majorana mass term:
\begin{equation}
{\cal{L}}_{\rm mass} = -\frac{1}{2} (M_{\rm R})_{ij}
\left[\nu^T_{{\rm R}i}\nu_{{\rm R}j}\right] - (M_D)_{ij}[\bar\nu_{{\rm L}i}
\nu_{{\rm R}j}] + \hbox{h.c.}~,
\end{equation}
with the Dirac mass matrix $M_D$ being proportional to the doublet Higgs
VEV.\footnote{Naively, one would expect $M_D$ to be similar to the
mass matrix $M_\ell$ for the charged leptons.}  If the eigenvalues of 
$M_R$ are much greater than those of $M_D$, then the neutrino mass matrix
\begin{equation}
{\cal{M}} = \left(
\begin{array}{cc}
0 & M_{\rm D} \\ M_{\rm D} & M_{\rm R}
\end{array} \right)
\end{equation}
has a set of large eigenvalues, corresponding to the eigenvalues of
$M_R$, and a set of extremely small eigenvalues, associated with the matrix
$M_D^2/M_R$.  This is the famous see-saw mechanism (\cite{YGRS}). 
As a result, one ends
up with a spectrum of neutrinos with both superheavy states and superlight
states:
\begin{equation}
{\cal{L}}_{\rm mass} \simeq -\frac{1}{2}
\left[\frac{M_{\rm D}^2}{M_{\rm R}}\right]_{ij} \bar\eta_{1i}\eta_{1j} -
\frac{1}{2} [M_{\rm R}]_{ij}\bar\eta_{2i}\eta_{2j}~.
\end{equation}
The light neutrinos $\eta_{1i}$ are mostly left-handed, while the heavy
neutrinos $\eta_{2i}$ are mostly right-handed.

The mass mixing discussed above, has a counterpart in the interactions
of the Majoron.  Although the field $\chi$ mostly couples to the heavy
fields $\eta_{2i}$, there will also be a small coupling of $\chi$ to $\eta_{1i}$.
That is, the Majoron $\chi$ as a result of the neutrino mass mixing
actually has also a small coupling to the ordinary left-handed neutrinos.
Specifically (\cite{CMP}), one finds
\begin{equation}
{\cal{L}}^{\rm int}_{\rm Majoron} = -\frac{h_{ij}}{2}
\bar\eta_{2i}i\gamma_5\eta_{2j}\chi - 
\left(\frac{hM_{\rm D}^2}{M^2_{\rm R}}\right)_{ij}
\bar\eta_{1i}i\gamma_5\eta_{2j}\chi~.
\end{equation}
It follows that the Majoron coupling to the light neutrinos is of order
$M_{\rm D}^2/M_{\rm R}^2\sim m_\nu/M_{\rm R}$, where $m_\nu$ is the mass
(matrix) for the light neutrinos.  The Majoron has an even weaker coupling
to ordinary matter, which is induced at one-loop order via mixing of the
$\chi$ with the $Z^o$.  One finds (\cite{CMP})
\begin{equation}
{\cal{L}}^{\rm eff}_{\rm matter} = i\frac{m_f}{v_\chi} \bar f\gamma_5
f\chi
\end{equation}
with the scale $v_\chi$ of order $v_\chi \sim (G_Fm_\nu)^{-1} \gg
{\rm TeV}$.  So clearly the Majoron in this model easily satisfy the
constraints imposed on additional dipole-dipole interactions in matter (\cite{FS}).

If Majorons exist, it is possible for the heaviest of the light neutrinos
to decay into the other neutrinos by Majoron emission.  The process
$\nu_i\to\nu_j\chi$, if it were fast enough, would serve to open up a
region of neutrino masses forbidden by cosmology.  For stable neutrinos,
one knows that neutrinos in the mass range from a few eV to a few 
GeV (\cite{KT}) overclose the Universe.  However, these bounds cease to apply
for unstable neutrinos.  If the lifetime $\tau$ for the neutrino decay
$\nu_i\to\nu_j\chi$ is much shorter than the Universe's lifetime
$T_o$, then effectively one can redshift the $\nu_i$ energy beyond its
mass $m_{\nu_i}$.  Hence the contribution of these neutrinos to the
energy density of the Universe is reduced to (\cite{CMP2})
\begin{equation}
\rho_{\nu_i} \sim m_{\nu_i}\left[\frac{\tau}{T_o}\right]^{1/2}
T_\nu^3
\end{equation}
where $T_\nu$ is the neutrino temperature now, $T_\nu \sim T_\gamma \sim
3^o~{\rm K}$.

The lifetime $\tau$ for the process $\nu_i\to \nu_j\chi$ naively scales
as  (\cite{CMP2})
\begin{equation}
\tau(\nu_i\to\nu_j\chi)\sim\frac{1}{m_{\nu_i}}
\left[\frac{M_{\rm R}}{M_{\rm D}}\right]^4
\end{equation}
and can be made short enough if $M_{\rm R}$ is not too large.  However,
in the simplest Majoron model
discussed here (\cite{CMP}) this lifetime is lengthened by
a further factor of $\left[\frac{M_{\rm R}}{M_{\rm D}}\right]^4$ (\cite{SV}),
making it very doubtful that $\tau < T_o$.  More elaborate models (\cite{Grevien}) restore the simple formula (247) and the possibility
that, through Majoron decay, neutrinos with masses in the ``forbidden"
cosmological range could exist.  This is not entirely an academic
exercise, as the existing bounds on $m_{\nu_\mu}$ and $m_{\nu_\tau}~
[m_{\nu_\mu} \leq 170~{\rm keV};~m_{\nu_\tau} < 24~{\rm MeV}$ (\cite{PDG})]
allow these particles to have masses precisely in this range.

\subsection{Global Symmetries and Gravity.}

In the preceding subsections I have discussed various interesting global
symmetries, which may be associated with the interactions of the Standard
Model, and have explored a bit the consequences of these symmetries.  There are,
however, some arguments one can adduce from the analysis of gravitational
interactions which bring into question the whole notion of having theories with exact global
symmetries.  I want to end my lectures by discussing this point briefly.

Perhaps the simplest way to see why gravitational interactions may cause
trouble is to focus on the ``No Hair" theorem for black holes.  Basically this
theorem (see, for example, \cite{bald}) asserts that black holes can be characterized only by a
few fundamental quantities, like mass and spin, but possess otherwise no other
quantum numbers.  Because black holes can absorb particles which carry global
charge, while carrying no global charge themselves, it appears that through
these processes one can get an explicit violation of whatever symmetry is
associated with the global charge.  That is, global charge can be lost when
particles carrying this charge are swallowed by a black hole.

One can parametrize the effect of the breaking of global symmetries by
gravitational interactions by adding to the low-energy Lagrangian 
non-renormalizable terms, scaled by inverse powers of the Planck mass
$M_{\rm P} \sim 10^{19}~{\rm GeV}$.  These terms, of course, should be
constructed so as to explicitly violate the symmetries in question.  Schematically, therefore, the full Lagrangian of the theory, besides containing
the usual Standard Model terms, should also include some effective 
non-renormalizable interactions containing various operators $O_n$, breaking
explicitly the Standard Model global symmetries:
\begin{equation}
{\cal{L}}^{\rm eff}_{\rm grav.~int.} = \sum_m \frac{1}{M_{\rm P}^n}~
O_n~.
\end{equation}
Here the dimension of the operators $O_n$, which explicitly breaks some of
the Standard Model global symmetries, is $n+4$.

Let me make two remarks.  First, the Lagrangian (248) can often be augmented
by other effective interactions which themselves break certain global
symmetries even more strongly than gravity.  For instance, an explicit mass
term for the right-handed neutrinos [cf Eq. (237)] does break $L$ directly
and more strongly than the operators in Eq. (248) do.  This said, however,
in what follows I will concentrate only on the gravitational effects 
embodied in Eq. (248).

Because $M_{\rm P} \sim 10^{19}~{\rm GeV} \gg (\sqrt{2}~G_F)^{-1/2} \sim
250~{\rm GeV}$, the naive expectation is that Eq. (248) cannot be that
important, except at superheavy scales.  This turns out to be true for the
interactions themselves, but fails when one considers the effect of Eq. 
(248) on the Nambu-Goldstone sector.  To demonstrate the first point, let me
consider the example of (B+L)-violation.  The dominant, $d=6$, (B+L)-violating
interaction induced by gravity schematically has the form (\cite{WWZ})
\begin{equation}
{\cal{L}}_{(B+L)-violation} \sim \frac{1}{(M_{\rm P})^2} 
u_i^cd_ju^c_kef_{ijk}~.
\end{equation}
Such a term leads to a proton lifetime, for the process
$p\to e^+\pi^0$, of order
\begin{equation}
\tau(p\to e^+\pi^0) \sim (M_{\rm P})^4 \sim
10^{46}~{\rm years}~,
\end{equation}
much greater than the present experimental bound  on this process
discussed earlier [Eq. (155)].  So the breaking of B+L provided through
gravitational effects is indeed irrelevant.

The situation is, however, different when one considers the Nambu-Goldstone
sector.  Let us consider again the simple example of spontaneously broken
Lepton number with its associated Majoron.  To the Lepton number
conserving potential, which forces the $SU(2)\times U(1)$ singlet field
$\sigma$ to acquire a VEV, one must now add non-renormalizable Lepton number
violating terms induced by the gravitational interactions.  The simplest
such term involves a dimension 5 operator.  Thus, one is invited to study
the potential
\begin{equation}
V_{\rm total} = \lambda\left(\sigma^{\dagger}\sigma -
\frac{V^2}{2}\right)^2 - \frac{\lambda^\prime}{M_{\rm P}}
(\sigma^{\dagger}\sigma)^2[\sigma + \sigma^{\dagger}]~.
\end{equation}
The first term above is clearly invariant under the Lepton number transformation
$\sigma\to e^{-2ia}\sigma$.  This is not so for the term which scales as
$M_{\rm P}^{-1}$.  Writing, as before,
\begin{equation}
\sigma \simeq \frac{V}{\sqrt{2}} \exp\left[i\frac{\chi}{V}\right]~,
\end{equation}
one sees that the effect of including the gravitational corrections is to
produce a mass term for the erstwhile Nambu-Goldstone field $\chi$.  One
finds, for the Majoron, a mass
\begin{equation}
m^2_\chi = \frac{\lambda^\prime}{2\sqrt{2}}
V^2 \left(\frac{V}{M}\right)~.
\end{equation}
Note that the size of the Majoron mass depends on the value of $V$, the scale
of the spontaneous breakdown of Lepton number.  For example, if we took
$V\sim {\rm TeV}$---the lowest it can be according to the bound of
Eq. (236)---then $m_\chi\sim V(V/M_{\rm P})^{1/2} \sim
10^{-8}~{\rm V}\simeq 10~{\rm KeV}$.  If $V$ is larger, the mass of the Majoron
grows as
\begin{equation}
m_\chi \sim 10\left[\frac{V}{\rm TeV}\right]^{3/2}~{\rm KeV}~.
\end{equation}
Clearly, if the Majoron is massive, some of its physical properties are
altered substantially.  For instance, it could happen that the decay
$\nu_i\to\nu_j\chi$ is actually kinematically forbidden!  Of course, the above
results are predicated on the assumption that the global Lepton number
symmetry is violated explicitly by a dim 5 interaction.  If the violation
were due to a higher dimensional operator of dimension $d$, then one finds
for the Majoron mass the formula
\begin{equation}
m_\chi \sim V\left(\frac{V}{M_{\rm P}}\right)^{\frac{d-4}{2}}
\end{equation}
which leads to masses which become smaller the larger $d$ is.

These considerations are particularly troubling for the $U(1)_{\rm PQ}$
solution to the strong CP problem (\cite{axiongrav}).  Not only potentially do
gravitational effects give an additional contribution to the axion mass,
but they can also alter the QCD potential so that $\bar\theta$ does not
finally adjust to zero!  One can understand what is going on by
schematically sketching the form of the effective axion potential in the
absence and in the presence of the $U(1)_{\rm PQ}$ breaking gravitational
interactions (\cite{Barr}).  Without gravity, a useful parametrization for the
physical axion effective potential, which follows from examining the
contributions of instantons (\cite{PQ}), is
\begin{equation}
V_{\rm axion} = -\Lambda^4_{\rm QCD} \cos a_{\rm phys}/f~.
\end{equation}
This potential displays the necessary periodicity in 
$a_{\rm phys}/f$, has a minimum
at $\langle a_{\rm phys}\rangle = \bar\theta_{\rm eff}=0$, and leads to an axion
mass $m_a = \Lambda_{\rm QCD}^2/f$.

Including gravitational effects changes the above potential by adding a
sequence of terms involving operators of different dimensions.  Let us just
consider one such term and examine the potential (\cite{Barr})
\begin{equation}
\tilde V_{\rm axion} = -\Lambda^4_{\rm QCD} \cos\frac{a_{\rm phys}}{f} -
\frac{cf^d}{M_{\rm P}^{d-4}}\cos
\left[\frac{a_{\rm phys}}{f}-\delta\right]~.
\end{equation}
Here $c$ is some dimensionless constant and $\delta$ is a CP-violating
phase which enters through the gravitational interactions.  This potential
modifies the formula for the axion mass, giving now
\begin{equation}
m_a^2 \simeq \frac{\Lambda^4_{\rm QCD}}{f^2} + c
\frac{f^{d-2}}{M_{\rm P}^{d-4}}~.
\end{equation}
For $f$ in the range of interest for invisible axions, the second term
above coming from the gravitational effects dominates the QCD mass
estimate for the axion, unless $c$ is extraordinarily small and/or
the dimension $d$ is rather large.  More troublesome still,
$\tilde V_{\rm axion}$ now no larger has a minimum at 
$\langle a_{\rm phys}\rangle = 0$.
Rather one finds a minimum of $\tilde V_{\rm axion}$ for values of
\begin{equation}
\bar\theta_{\rm eff} = \frac{\langle a_{\rm phys}\rangle}{f} \simeq
c\sin\delta \frac{f^d}{M_{\rm P}^{d-4}\Lambda^4_{\rm QCD}}~.
\end{equation}
That is, the gravitational effects (provided there is a CP violating phase
associated with them) induce a non-zero
$\bar\theta$, even in the presence of a $U(1)_{\rm PQ}$ symmetry!  To satisfy the bound $\bar\theta \leq 10^{-10}$ again
necessitates that $d$ be large and/or that the constant $c$ be extraordinarily
small.

To date there is no clear resolution to this problem and it could be that
these considerations actually vitiate the chiral solution to the strong
CP problem.  Since this is the most appealing solution to this conundrum,
this is somewhat troubling.  Nevertheless, it is worth noting a number of
points.  First, one does not really understand quantum gravity.  Thus it is
possible that when matters are better understood the effective global
symmetry breaking interactions we introduced may in fact not be there at all,
or be tremendously suppressed.  Second, there are some encouraging results in
this direction coming from string theory. Axions associated with broken
chiral symmetries arise very naturally in string 
theory (\cite{Witten}).  Furthermore,
CP is conserved, at least in higher dimensions in string theory (\cite{Kaplan}),
so perhaps it is possible that $\sin\delta=0$.  Finally, there are arguments
that for large compactification radii, the effective $U(1)_{PQ}$
symmetries are broken very little in strings, so that the tiny number needed
for $c~[c\leq 10^{-51}]$ may not be out of the question (see, for example, \cite{choi}).

Irrespective of the above considerations, one should note that if the
gravitational effects induce values of $\bar\theta < 10^{-10}$, so that the
strong CP problem is still solved by imposing a $U(1)_{PQ}$ symmetry, then
also the axion mass is approximately given by its QCD form.  Thus, perhaps
the best way to resolve these thorny theoretical questions is to find
experimental evidence for the existence of invisible axions, with the
canonical properties!
\vskip.3cm

{\flushleft{\bf Acknowledgments}}
\vskip.3cm

I am grateful to Professor W. Plessas for the very nice hospitality 
shown to me at
Schladming.  This work is supported in part by the Department of Energy
under Grant No. FG03-91ER40662, Task C.

\end{document}